\documentclass[apj] {emulateapj}
\usepackage{apjfonts}
\usepackage{epstopdf}
\usepackage{amsmath}
\usepackage{subfigure}
\def\wig#1{\mathrel{\hbox{\hbox to 0pt{%
          \lower.5ex\hbox{$\sim$}\hss}\raise.4ex\hbox{$#1$}}}}

\shorttitle{GJ 436b}

\newcommand{\me}{$M_{\oplus}$}

\newcommand{\teff}{$T_{\rm eff}$}

\newcommand{\cp}{\citep}
\newcommand{\ct}{\citet}

\newcommand{\fsed}{$f_{\rm sed}$} 
\newcommand{\fhaze}{$f_{\rm haze}$} 
\newcommand{\nas}{Na$_2$S}
\newcommand{\rchi}{$\chi_{\rm red}^2$}
\newcommand{\kzz}{$K_{zz}$}
\newcommand{\jwst}{\emph{JWST}}
\newcommand{\tint}{ $T_{\rm int}$}

\slugcomment{Draft for ApJ}

\begin{document}

\title{Forward and Inverse Modeling of the Emission and Transmission Spectrum of GJ 436b:  Investigating Metal Enrichment, Tidal Heating, and Clouds }

\author{Caroline V. Morley\altaffilmark{1, 2}, Heather Knutson\altaffilmark{3},  Michael Line\altaffilmark{4}, Jonathan J. Fortney\altaffilmark{5}, Daniel Thorngren\altaffilmark{6}, Mark S. Marley\altaffilmark{7}, Dillon Teal\altaffilmark{5}, Roxana Lupu\altaffilmark{7}}

\altaffiltext{1}{Department of Astronomy, Harvard University, Cambridge, MA 02138; caroline.morley@cfa.harvard.edu}
\altaffiltext{2}{NASA Sagan Fellow}
\altaffiltext{3}{Division of Geological and Planetary Sciences, California Institute of Technology} 
\altaffiltext{4}{School of Earth and Space Exploration, Arizona State University} 
\altaffiltext{5}{Department of Astronomy and Astrophysics, University of California, Santa Cruz}
\altaffiltext{6}{Department of Physics, University of California, Santa Cruz}
\altaffiltext{7}{NASA Ames Research Center} 

\begin{abstract}
 
The Neptune-mass GJ 436b is one of the most-studied transiting exoplanets with repeated measurements of both its thermal emission and transmission spectra. We build on previous studies to answer outstanding questions about this planet, including its potentially high metallicity and tidal heating of its interior. We present new observations of GJ 436b's thermal emission at 3.6 and 4.5 \micron, which reduce uncertainties in estimates of GJ 436b's flux at those wavelengths and demonstrate consistency between \emph{Spitzer} observations spanning more than 7 years. We analyze the \emph{Spitzer} thermal emission photometry and \emph{Hubble} WFC3 transmission spectrum in tandem.  We use a powerful dual-pronged modeling approach, comparing these data to both self-consistent and retrieval models.  We vary the metallicity, intrinsic luminosity from tidal heating, disequilibrium chemistry, and heat redistribution. We also study the effect of clouds and photochemical hazes on the spectra, but do not find strong evidence for either. The self-consistent and retrieval modeling combine to suggest that GJ 436b has a high atmospheric metallicity, with best fits at or above several hundred times solar metallicity, tidal heating warming its interior with best-fit intrinsic effective effective temperatures around 300--350 K, and disequilibrium chemistry. High metal-enrichments ($>600\times$ solar) can only occur from the accretion of rocky, rather than icy, material. Assuming \tint$\sim$300--350 K, we find that $Q'\sim 2\times10^5$--$10^6$, larger than Neptune's $Q'$, and implying a long tidal circularization timescale for the planet's orbit.  We suggest that Neptune-mass planets may be a more diverse class than previously imagined, with metal-enhancements potentially spanning several orders of magnitude, to perhaps over 1000$\times$ solar metallicity. High fidelity observations with instruments like JWST will be critical for characterizing this diversity. 
\end{abstract}

 
\section{Introduction}

Determining the compositions of exoplanets ranging from Earth-mass to Jupiter-mass in different environments is a key goal of exoplanetary research. Planetary compositions are shaped by the details of planet formation and altered by atmospheric physics and chemistry. Over a decade after its discovery by \ct{Butler04}, GJ 436b remains the planet in its Neptune-mass class for which we have obtained the most detailed observations of its atmosphere. 

GJ 436b was discovered to transit by \ct{Gillon07a} and, as the smallest transiting planet in 2007 and a favorable target for observations, immediately became a target for atmospheric characterization studies with the \emph{Spitzer Space Telescope} and \emph{Hubble Space Telescope}. It remains one of the most favorable and interesting targets for followup spectroscopic studies: to date a total of 18 secondary eclipses and 8 transits have been observed with \emph{Spitzer}, along with 7 transits with \emph{HST} \cp{Deming07, Demory07, Gillon07b, Stevenson10, Beaulieu11, Knutson11, Knutson14a}. 

The atmosphere of GJ 436b has been a perennial challenge to understand. Previous observations and modeling efforts, which we describe below, have suggested high metallicity compositions with strong vertical mixing. Many of these conclusions rest on the robustness of the \emph{Spitzer} 3.6 and 4.5 \micron\ eclipses. Here, we move forward to study this planet using both its thermal emission photometry and its transmission spectrum, adding three new eclipse observations at these two wavelengths and analyzing the dataset with a powerful dual-pronged approach of self-consistent and retrieval modeling.

\subsection{Observations and Interpretation of Thermal Emission}
Secondary eclipse measurements allow us to infer the planet's brightness, and therefore temperature, as a function of wavelength when the planet passes behind the host star. A planet will appear fainter, and therefore create a shallower occultation, at wavelengths of strong absorption features, and appear brighter at wavelengths of emission features. 

The first secondary eclipse measurements of GJ 436b were observed at 8 $\micron$, while \emph{Spitzer} was still operating cryogenically \cp{Deming07, Demory07}. These observations revealed that GJ 436b has a high eccentricity, $\sim$0.15, which, given predicted tidal circularization timescales, suggests the presence of a companion and of potential tidal heating \cp{Ribas08, Batygin09}. 

With an equilibrium temperature around 700--800 K, GJ 436b is cool enough that models assuming thermochemical equilibrium predict high CH$_4$ abundance and low CO and CO$_2$ abundance, which would result in a deeper occultation at 4.5 \micron\ than 3.6 \micron. However, when \ct{Stevenson10} published the first multi-wavelength thermal emission spectrum of GJ 436b, measuring photometric points at 3.6, 4.5, 5.8, 8.0, 16, and 24 \micron, they found that its occultation was deeper at 3.6 \micron\ and shallower  at 4.5 \micron\ and suggested methane depletion due to photo-dissociation as an explanation. Additional studies have reanalyzed these observations and observed additional secondary eclipses \cp{Knutson11, Lanotte14}. In particular, the analysis by \ct{Lanotte14} revealed a significantly shallower 3.6 \micron\ eclipse and somewhat shallower 8.0 \micron\ eclipse; no detailed atmospheric studies have been carried out since these revisions. 

From the time of the initial observations of GJ 436b's thermal emission, it has been a major challenge to find self-consistent models that adequately explain the data. \ct{Madhu11d} found using retrieval algorithms that the atmosphere is best fit by an atmosphere rich in CO and CO$_2$ and depleted in CH$_4$. \ct{Line11} used disequilibrium chemical models including the effect of photochemistry, but found that they were not able to reproduce the low observed methane abundance. \ct{Moses13} found that high metallicities (230--1000$\times$ solar) favor the high CO and CO$_2$ abundances inferred from the observations. \ct{Agundez14}, noting the high eccentricity of GJ 436b, studied the effect of tidal heating deep in the atmosphere on the chemistry and find that significant tidal heating and high metallicities fit the observed photometry best.

\subsection{Observations and Interpretation of Transmission Spectrum}

Wavelength-dependent observations of the transit depth of GJ 436b allow us to probe the composition of GJ 436b's day--night terminator. At wavelengths with strong absorption features, the planet will occult a larger area of the star, resulting in a deeper transit depth. 
\ct{Pont09} observed the transmission spectrum of GJ 436b from 1.1 to 1.9 \micron\ with NICMOS on \emph{HST} but due to systematic effects were unable to achieve high enough precision to detect the predicted water vapor feature. \ct{Beaulieu11} presented transit measurements in \emph{Spitzer}'s 3.6, 4.5, and 8.0 \micron\ filters that showed higher transit depths at 3.6 and 8.0 \micron\ than at 4.5 \micron, indicating strong methane absorption. \ct{Knutson11} analyzed the same data and suggested that variable stellar activity caused the observed transit depth at 3.6 \micron\ to vary between epochs. However, these data were reanalyzed again by \ct{Lanotte14} and \ct{Morello15} with new techniques, which both found that the transit depths were constant in the different bandpasses and remained constant between epochs of observations. 

More recently, \ct{Knutson14a} used WFC3 on \emph{HST} to measure the transmission spectrum from 1.1--1.7 \micron. Like \ct{Pont09}, they do not detect a water vapor feature, but with their higher S/N spectrum are able to rule out a cloud-free H/He-dominated atmosphere to high confidence (48$\sigma$). The spectrum is consistent with a high cloud at pressures of $\sim$1 mbar, or a H/He poor (3\% H/He by mass, 1900$\times$ solar) atmospheric composition.

\subsection{A Third Body in the GJ 436 System?} \label{thirdbody}

Because the tidal circularization timescales are predicted to be shorter than the age of the star, the non-zero eccentricity of GJ 436b has suggested that it may have at least one companion in the system; however, a number of searches for additional planets did not find additional bodies \cp{Demory07, Deming07, Maness07, Alonso08, Ribas08, Caceres09, Ballard10a, Ballard10b, Beust12}. \ct{Stevenson12} announced the detection of two candidate sub-Earth-size planets in the system, but later work by \ct{Lanotte14} did not find any evidence of these candidate companions.

\subsection{The Need for an Additional Atmospheric Study}

Here, we build on this extensive history of both observations and modeling for this enigmatic warm Neptune to answer the still-outstanding questions about this planet.  Do the revisions in the eclipse points from \ct{Lanotte14} change the inferred composition? Is it truly ultra-high ($>$300$\times$ solar) metallicity? What atmospheric physics must be present for a Neptune-mass planet to have the observed spectra and inferred atmospheric composition? 

To these ends, we present an additional three secondary eclipse observations (1 at 3.6 \micron, 2 at 4.5 \micron), demonstrating the robustness of these observations with modern \emph{Spitzer} observational and analysis techniques. We study both the thermal emission and transmission spectra of GJ 436b in tandem, including the published dataset of \emph{Spitzer} photometry spanning from 3.6 to 16 \micron\ and the transmission spectrum from \emph{HST}/WFC3. Unlike most previous studies, we investigate whether including clouds or hazes in GJ 436b's atmosphere can match both sets of observations for Neptune-like compositions (50--300 $\times$ solar), without invoking ultra-high metallicity (>1000$\times$ solar) compositions. We combine our self-consistent treatment with results from chemically-consistent retrievals that do not include clouds, and show that H/He-poor atmospheric compositions with tidal heating provide the most precise fit to GJ 436b's thermal emission spectrum, while also fitting the transmission spectrum. 

\subsubsection{Format of this work}

In Section \ref{dataanalysis} we describe the observations and data analysis. In Section \ref{modeling}, we describe the modeling tools used to simulate the observations, including both self-consistent and retrieval models. In Section \ref{selfconsistent} we compare the data to self-consistent models; in Section \ref{retrievals} we use retrieval algorithms to retrieve chemical abundances and pressure--temperature profile and compare these results with the results from self-consistent modeling.

\section{Observations and Data Analysis} \label{dataanalysis}

\begin{deluxetable*}{lccccccccc}
\tabletypesize{\scriptsize}
\tablecaption{\textit{Spitzer} Observation Details\label{obs_param}}
\tablewidth{0pt}
\tablehead{
\colhead{$\lambda$ ($\mu m$)} & \colhead{UT Start Date} & \colhead{Length (h)} & \colhead{$n_{img}$\tablenotemark{a}} & \colhead{$t_{int}$ (s)\tablenotemark{b}} & \colhead{$t_{trim} $\tablenotemark{c}} & \colhead{$n_{bin} $\tablenotemark{c}} & \colhead{$r_{pos}$\tablenotemark{c}} & \colhead{$r_{phot}$\tablenotemark{c}} & \colhead{Bkd ($\%$)\tablenotemark{e}}}
\startdata
3.6 & 2008-01-30 & 5.9 & 163,200 & 0.1 & 1.0 & 192 & 3.0 & 2.8 & 0.05\\
3.6 & 2014-07-29 & 4.5 & 122,112 & 0.1 & 1.0 & 128 & 2.0 & 2.5 & 0.25\\
4.5 & 2008-02-02 & 5.9 & 49,920 & 0.4 & 3.0 & 32 & 2.0 & 2.9 & 0.09\\
4.5 & 2011-01-24 & 6.1 & 51,712 & 0.4 & 2.0 & 32 & 2.0 & 4.5 & 0.38\\
4.5 & 2014-08-11 & 4.5 & 122,112 & 0.1 & 0.5 & 128 & 2.0 & 2.7 & 0.11\\
4.5 & 2015-02-25 & 4.5 & 122,112 & 0.1 & 0.5 & 128 & 2.0 & 2.8 & 0.12
\enddata
\tablenotetext{a}{Total number of images.}
\tablenotetext{b}{Integration time.}
\tablenotetext{c}{$t_{trim}$ is the amount of time in hours trimmed from the start of each time series, $n_{bin}$ is the bin size used in the photometric fits, $r_{pos}$ is the radius of the aperture used to determine the position of the star on the array, and $r_{phot}$ is the radius of the photometric aperture in pixels.}
\tablenotetext{e}{Sky background contribution to the total flux for the selected aperture.}
\end{deluxetable*}

\subsection{Photometry and Instrumental Model}

These observations were obtained in the 3.6 and 4.5 $\mu$m bandpasses using the Infra-Red Array Camera (IRAC) on the \textit{Spitzer Space Telescope}.  In this paper we present three new secondary eclipse observations of this planet, including a 3.6~\micron~observation obtained on UT 2014 Jul 29 and two 4.5~\micron~observations obtained on UT 2014 Aug 11 and UT 2015 Feb 25, respectively, as part of \emph{Spitzer} program 50056 (PI: Knutson).  We also re-examine three archival eclipse observations including a 3.6~\micron~eclipse from UT 2008 Jan 30, as well as 4.5~\micron~eclipses from UT 2008 Feb 2 and UT 2011 Jan 24 \citep{Stevenson10,Stevenson12,Lanotte14}.  Eclipses from 2008 were observed during \textit{Spitzer's} cryogenic mission, while the remaining eclipses were observed during the extended warm mission.  All eclipses were observed in subarray mode, with integration times and observation durations given in Table \ref{obs_param}.  Our new 2014-2015 observations included a now-standard 30-minute peak-up pointing observation prior to the start of our science observations.  This adjustment corrects the initial telescope pointing in order to place the star near the center of the pixel where the effect of intrapixel sensitivity variations is minimized.  

We utilize BCD image files for our photometric analysis and extract BJD$_{\rm UTC}$ mid-exposure times using the information in the image headers.  The sky background for each 32$\times$32 pixel image is calculated by excluding a circular region with a radius of twelve pixels centered on the star, taking all of the remaining pixels and trimming outliers greater than three standard deviations away from the median, and then fitting a Gaussian function to a histogram of the remaining pixels. We calculate the flux-weighted centroid position of the star on the array and derive the corresponding total flux in a circular aperture for each individual image as described in previous studies \citep[e.g.][]{Lewis13,Deming15,Kammer15}. We consider both fixed and time varying photometric aperture sizes in our fits but find that in all cases we obtain a lower RMS and reduced levels of time-correlated (``red") noise in our best-fit residuals using fixed apertures, in good agreement with the conclusions of \citet{Lanotte14}.  We consider apertures with radii ranging between $2.0-5.0$ pixels, where we step in increments of 0.1 pixels between $2.0-3.0$ pixels and in $0.5$ pixel increments for larger radii.  

\begin{figure}[h]
\includegraphics[width=3.4in]{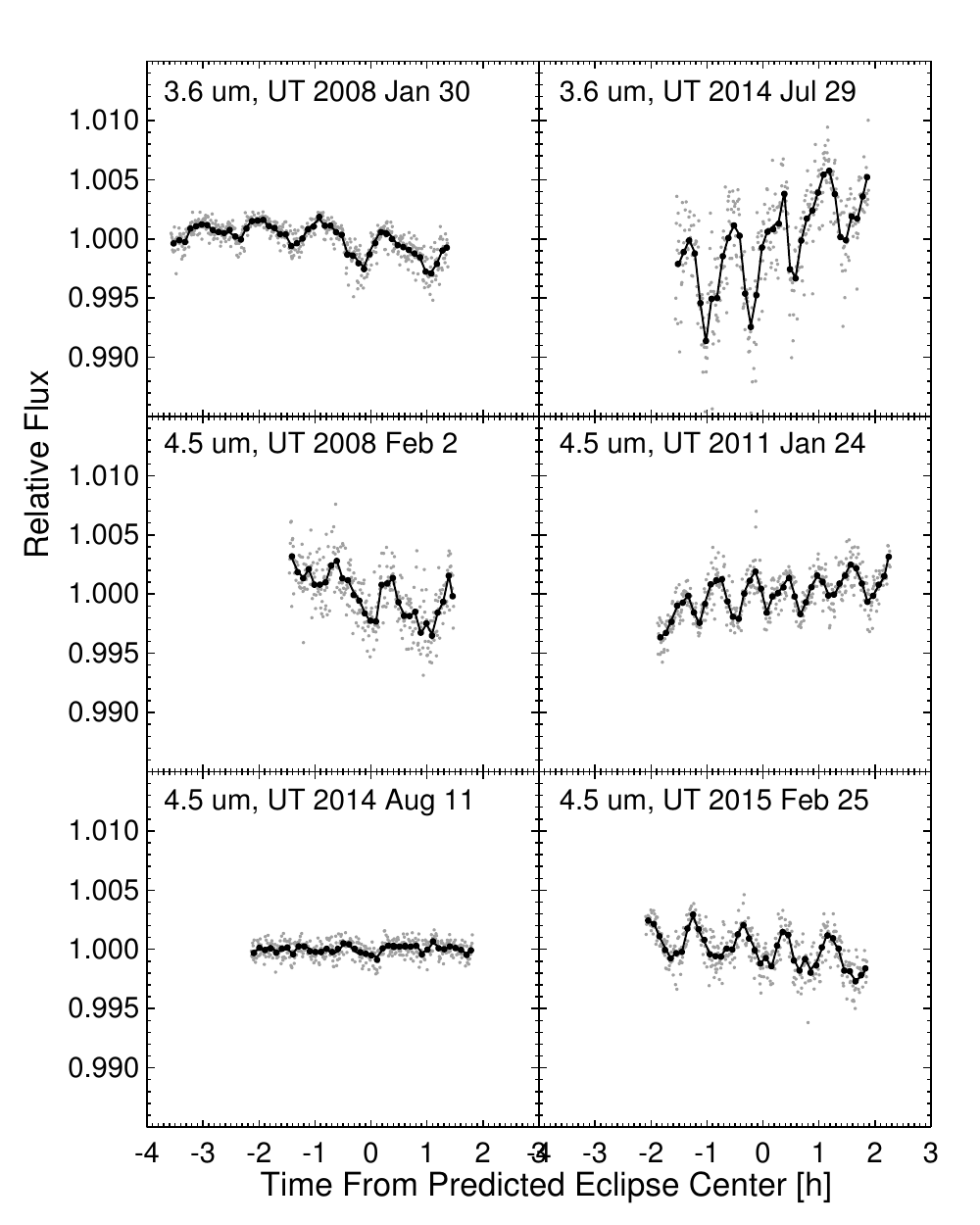}
\caption{Raw \textit{Spitzer} 3.6 and 4.5 $\mu$m photometry as a function of time from the center of eclipse phase reported in \citet{Knutson11}.  We bin the photometry in 30~s (grey filled circles) and 5 minute (black filled circles) intervals, and overplot the best fit instrumental models binned in 5 minute intervals for comparison (solid lines).}
\label{rawfig}
\end{figure}

\begin{figure}[h]
\includegraphics[width=3.4in]{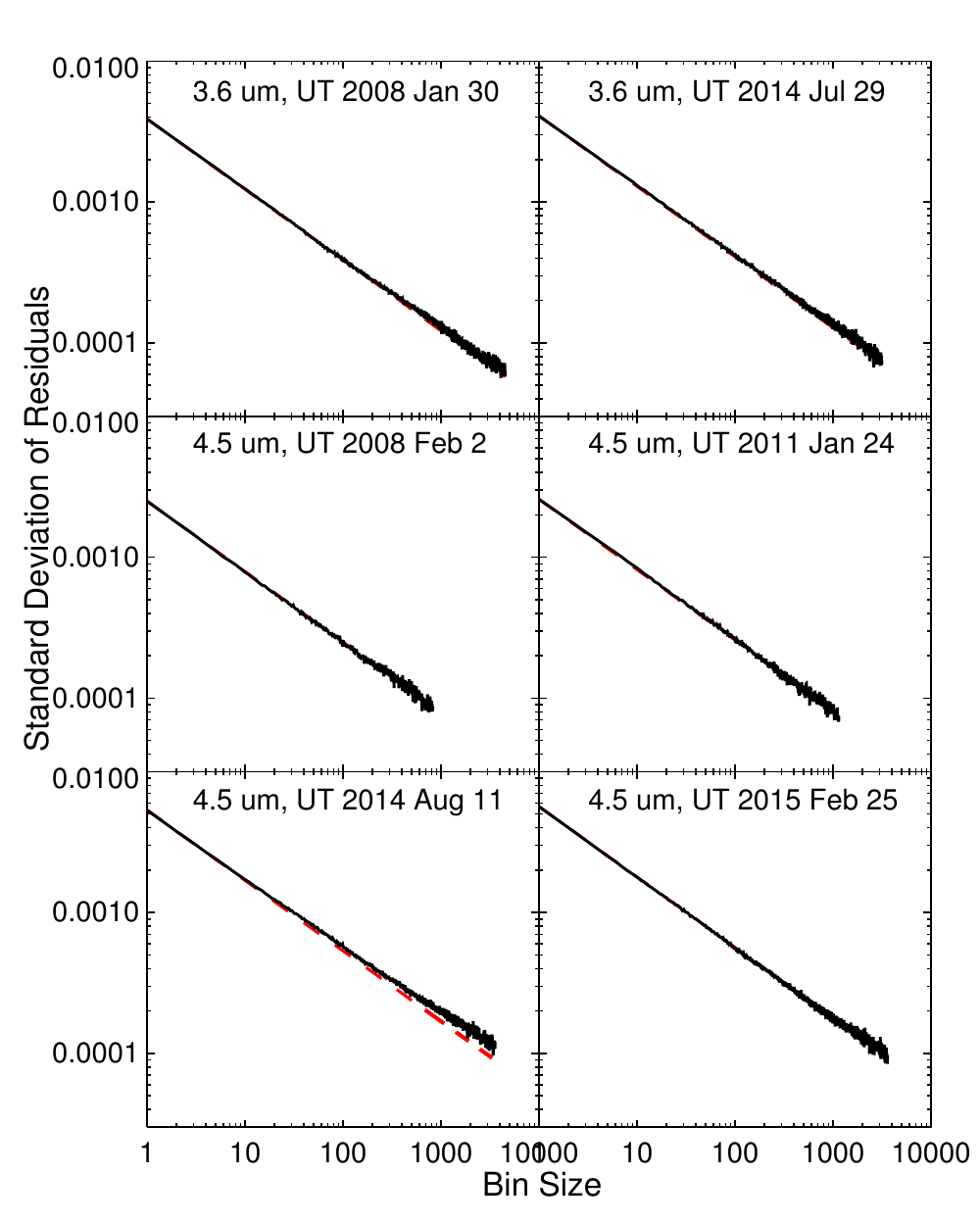}
\caption{Standard deviation of the best-fit residuals as a function of the number of data points per bin} (black lines).  We over plot the expected $1/\sqrt n$ scaling for Gaussian noise as red dashed lines, where we have normalized these lines to match the standard deviation of the unbinned residuals.
\label{rootn_fig}
\end{figure}

The sensitivity of individual 3.6 and 4.5~\micron~IRAC pixels varies from the center to the edge; when combined with short-term telescope pointing oscillations, this produces variations in the raw stellar fluxes plotted in Fig. \ref{rawfig}.  We correct for this effect using the pixel-level decorrelation (PLD) method \citep{Deming15}, which produces results that are comparable to or superior to those from a simple polynomial decorrelation or pixel mapping method for light curves with durations of less than ten hours \citep[for a discussion of the PLD method applied to longer phase curve observations, see][]{Wong15}.  We utilize the raw flux values in a $3\times3$ grid of pixels centered on the position of the star, and then normalize these individual pixel values by dividing by the total flux in each $3\times3$ postage stamp.  We then incorporate these light curves into an instrumental model given by: 

\begin{equation}
F_{model}(t) = \frac{\sum\limits_i w_i F_i (t)}{\sum\limits_i F_i (t)} 
\end{equation}
where $F_{model}$ is the predicted stellar flux in an individual image, $F_i$ is the measured flux in the $i^{th}$ individual pixel, and $w_i$ is the weight associated with that pixel.  We leave these weights as free parameters in our fit, and solve for the values that best match our observed light curves simultaneously with our eclipse fits.  

\begin{figure}[h]
\epsscale{1.2}
\includegraphics[width=3.4in]{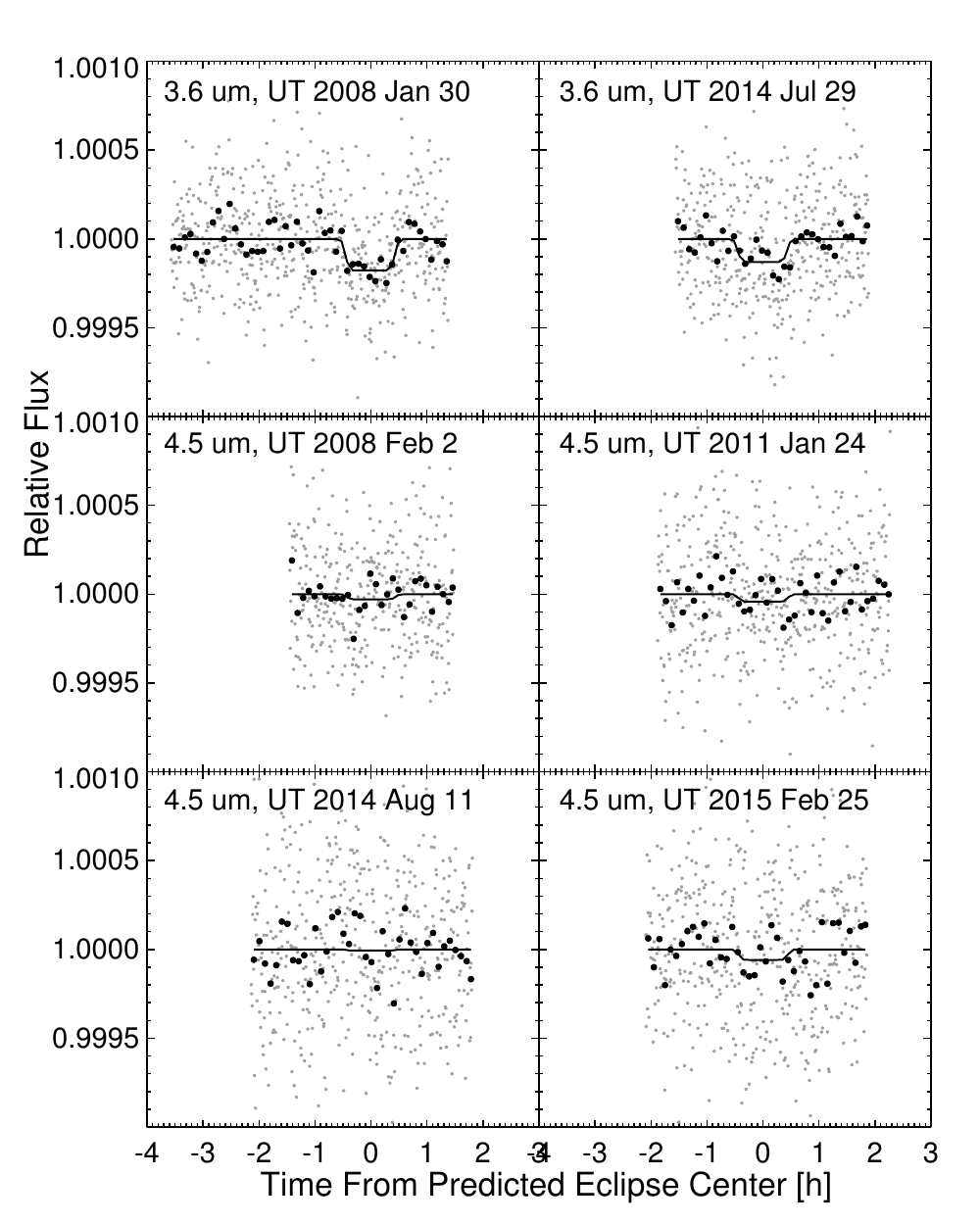}
\caption{Normalized \textit{Spitzer} 3.6 and 4.5 $\mu$m light curves as a function of time from the predicted center of eclipse, where we have divided out the best-fit instrumental model shown in Fig. \ref{rawfig}. The normalized flux is binned in 30 s (grey filled circles) and 5 minute (black filled circles) intervals, and best fit eclipse model light curves are over plotted for comparison (solid lines).}
\label{normfig}
\end{figure}

Following the example of \citet{Deming15}, we fit this model to binned light curves with optimal bin sizes given in Table \ref{obs_param}, where we considered a range of bin sizes between 1--512 points sampled in steps of 2$^n$.  After identifying the best-fit model we apply this solution to the unbinned light curves in order to generate a corresponding vector of unbinned residuals, which we use to evaluate the noise properties of the data. As discussed in \cite{Deming15} and \cite{Kammer15}, we create a metric to measure the noise properties of a given version of the photometry by calculating the root mean square (RMS) variance of the residuals as a function of bin size (Fig. \ref{rootn_fig}).  We then take the difference between a Gaussian noise model with $1/\sqrt n$ scaling and the observed RMS as a function of bin size, square the difference, and sum over all bins.   We then pick the version of the photometry that has the lowest amount of red noise as measured by our least squares metric after discarding solutions where the RMS of the best-fit residuals is more than 1.1 times higher than the lowest RMS version of the photometry.  We trim a small section of data from the start of each light curve in order to remove the exponential ramp, which is another well-known feature of the IRAC 3.6 and 4.5~\micron~arrays \citep[e.g.,][]{Lewis13,Zellem14}.  After selecting the optimal aperture and bin sizes, we examine the normalized light curves after detector effects have been removed and vary the trim duration until we identify the value which minimizes the red noise in the best-fit residuals, in agreement with our criteria for selecting the optimal aperture and bin size. Because we do not include an exponential function as part of our model fit, this criteria serves to identify trim durations which minimize the presence of the exponential ramp in our data..  We then re-run our previous analysis in order to ensure that our aperture and bin sizes are still optimal given this new trim duration.  

\begin{deluxetable*}{lccccccc}
\tabletypesize{\scriptsize}
\tablecaption{Best Fit Eclipse Parameters\label{eclipse_param}}
\tablecolumns{5}
\tablewidth{0pt}
\tablehead{
\colhead{$\lambda$ ($\mu m$)} & \colhead{UT Start Date}  & \colhead{$F_p/F_*$ (ppm)} & \colhead{$F_p/F_{*,avg}$ (ppm)\tablenotemark{a}} & \colhead{T$_{\rm bright}$ (K)\tablenotemark{a}} & \colhead{$T_s$\tablenotemark{b}} & $O-C$ (d)\tablenotemark{c} }
\startdata
3.6 & 2008-01-30 & $177\pm31$ & $151\pm27$ & $879^{+29}_{-27}$ & $4496.4888\pm0.0012$ & $-0.0007\pm0.0012$ \\
 3.6 & 2014-07-29 & $133\pm35$ &  & & $6868.0655\pm0.0054$ & $-0.0006\pm0.0054$ \\ 
 4.5 & 2008-02-02 & $30\pm36$ & $29\pm20$ & $<633$ & $4499.1334$\tablenotemark{d}\phantom{$pm0.000$} & \\ 
 4.5 & 2011-01-24 & $37\pm37$ & & & $5585.7756$\tablenotemark{d}\phantom{$pm0.000$} & \\
 4.5 & 2014-08-11 & $64\pm45$ & & & $6881.2856$\tablenotemark{d}\phantom{$pm0.000$} & \\ 
 4.5 & 2015-02-25 & $1\pm44$ & & & $7079.5779$\tablenotemark{d}\phantom{$pm0.000$} & 
\enddata
\tablenotetext{a}{We report the error-weighted mean eclipse depths at 3.6 and 4.5~\micron.  Brightness temperatures are calculated using a PHOENIX stellar model interpolated to match the published stellar temperature and surface gravity from \citet{vonbraun12}.}
\tablenotetext{b}{BJD$_{\rm UTC}$ - 2,450,000.}
\tablenotetext{c}{Observed minus calculated eclipse times, where we have accounted for the uncertainties in both the measured and predicted eclipse times as well as the XX s light travel time delay in the system.  We calculate the predicted eclipse time using the best-fit eclipse orbital phase from \citet{Knutson11}.}
\tablenotetext{d}{We allow the eclipse times in this bandpass to vary as free parameters in our fit, but we use the orbital phase and corresponding uncertainty from \citet{Knutson11} as a prior constraint in the fit.}
\tablenotetext{e}{$2\sigma$ upper limit based on the error-weighted average of the four 4.5~\micron~eclipse measurements.}
\end{deluxetable*}

\subsection{Eclipse Model and Uncertainty Estimates}

We generate our secondary eclipse light curves using the routines from \citet{mandel02}, where we fix the planet-star radius ratio, orbital inclination, eccentricity $e$, longitude of periapse $\omega$, and ratio of the orbital semi-major axis to the stellar radius to their best-fit values from \citet{Lanotte14}.  We allow individual eclipse depths and center of eclipse times to vary as free parameters in our fits to the 3.6~\micron~data.  We find that the eclipse depth in individual $4.5$~\micron~observations is consistent with zero, and therefore place a Gaussian prior on the phase of the secondary eclipse in order to constrain the best-fit eclipse time.  We implement this prior as a penalty in $\chi^2$ proportional to the deviation from the error-weighted mean center-of-eclipse phase and corresponding uncertainty from \citet{Knutson11}.  Although we also calculate the best-fit eclipse orbital phase using the $e$ and $\omega$ values from \citet{Lanotte14} and find that it is consistent with the value from \citet{Knutson11}, the corresponding uncertainty is substantially larger than that reported in \citet{Knutson11}.  This is not surprising, as the measured times of secondary eclipse constrain $e \cos \omega$ while $e \sin \omega$ is typically derived from fits to radial velocity data and has larger uncertainties \citep[e.g.][]{Pal10,Knutson14c}.  The uncertainties in the values for $e$ and $\omega$ reported in \citet{Lanotte14} are therefore likely to be dominated by the $e \sin \omega$, while $e \cos \omega$ is well-measured from secondary eclipse photometry alone.  

We chose not to include a prior for our 3.6 micron fits because we wanted to derive an independent estimate of the eclipse center time and phase. However, to test the effect of this choice, we repeat our 3.6~\micron~fits including this prior on the eclipse phase and find that the measured eclipse depths change by less than $0.1$ sigma, as expected for cases where the eclipse is detected at a statistically significant level.

We fit our combined eclipse and instrumental noise model to each light curve using a Levenberg-Marquardt minimization routine with uniform priors on all parameters except the 4.5~\micron~eclipse time as described in the previous paragraph.  Our model includes nine pixel weight parameters, two eclipse parameters, and a linear function of time in order to account for long-term instrumental and stellar trends.  We show the resulting light curves and best-fit eclipse models after dividing out the best-fit instrumental noise model and linear function of time in Fig. \ref{normfig}.  Uncertainties on model parameters are calculated using a Markov chain Monte Carlo (MCMC) analysis with $10^6$ steps initialized at the location of the best-fit solution from our Levenberg-Marquardt minimization.  We trim any remaining burn-in at the start of the chain by checking to see where the $\chi^2$ value of the chain first drops below the median value over the entire chain, and trim all points prior to this step.  We find that in all cases our probability distributions for the best-fit eclipse depths and times are Gaussian and do not show any correlations with other model parameters.  We therefore take the symmetric 68\% interval around the median parameter value as our $1\sigma$ uncertainties.  


\section{Atmospheric Modeling} \label{modeling}

We use a combination of self-consistent modeling and retrieval algorithms to model the atmosphere of GJ 436b and match its spectrum. The self-consistent modeling mirrors that used in \ct{Morley15}; our suite of tools includes a 1D radiative--convective model to calculate the pressure--temperature structure, a photochemical model to calculate the formation of hydrocarbons that may form hazes, and a cloud model to calculate cloud mixing ratios, altitudes, and particle sizes. We calculate spectra in different geometries and wavelengths using a transmission spectrum model, a thermal emission spectrum model, and an albedo model. We also use a retrieval model, \texttt{CHIMERA} \cp{Line12, Line13a, Line14a} to explore the thermal emission spectrum. In the following subsections we will briefly discuss each of these calculations. 

We fit our models to the thermal emission and transmission spectra separately and then analyze the regions of parameter space where the same model parameters fit both the thermal emission and transmission spectra. 

\subsection{1D Radiative--Convective Model}

We calculate the temperature structures of GJ 436b's atmosphere assuming radiative--convective equilibrium. These models are more extensively described in \ct{Mckay89, Marley96, Burrows97, Marley99, Marley02, Fortney05, Saumon08, Fortney08b}. Our opacity database for gases is described in \ct{Freedman08, Freedman14}. We calculate the effect of cloud opacity using Mie theory, assuming spherical particles. Optical properties of sulfide and salt clouds and soot haze are from a variety of sources and presented in \ct{Morley12} and \ct{Morley13}. 

To calculate P--T profiles for models with greater than 50$\times$ solar metallicity, we make the same approximation as used in \ct{Morley15}. We multiply the total molecular gas opacity by a constant factor (e.g. we multiply the 50$\times$ solar opacities by 6 to approximate the opacity in a 300$\times$ solar composition atmosphere). We change the abundances of hydrogen and helium separately to calculate collision-induced absorption. This approximation is appropriate for the results explored here; for future work, e.g. comparing models to \emph{JWST} data, new k-coefficients at 100--1000× solar metallicity should be used.

\subsection{Equilibrium Chemistry}

After calculating the pressure--temperature profiles of models with greater than 50$\times$ solar metallicity, we calculate the gas abundances assuming chemical equilibrium along that profile. We use the Chemical Equilibrium with Applications model (CEA, Gordon \& McBride 1994) to compute the thermochemical equilibrium molecular mixing ratios (with applications to exoplanets see, \ct{Visscher10, Line10, Moses11, Line11, Line13b}).  CEA minimizes the Gibbs Free Energy with an elemental mass balance constraint given a local temperature, pressure, and elemental abundances. We include molecules containing H, C, O, N, S, P, He, Fe, Ti, V, Na, and K. We account for the depletion of oxygen due to enstatite condensation by removing 3.28 oxygen atoms per Si atom \cp{Burrows99}.   When adjusting the metallicity all elemental abundances are rescaled equally relative to H, ensuring that the elemental abundances sum to one.  

\subsection{Photochemical Haze Model}

We use results from photochemical modeling in \ct{Line11}. Briefly, the computations use the Caltech/JPL photochemical and kinetics model, KINETICS (a fully implicit, finite difference code), which solves the coupled continuity equations for each species and includes transport via both molecular and eddy diffusion \cp{Allen81, Yung84, Moses05}. We use results for 50$\times$ solar composition, \kzz=10$^8$ cm$^2$/s (Figures 5, 6 and 7 in \ct{Line11}).

We follow the approach developed in \ct{Morley13} and used for GJ 1214b in \ct{Morley15} to calculate the locations of soot particles based on the photochemistry. We sum the number densities of the five soot precursors (C$_2$H$_2$, C$_2$H$_4$, C$_2$H$_6$, C$_4$H$_2$, and HCN) to find the total mass in soot precursors. We assume that the soots form at the same altitudes as the soot precursors exist: we multiply the precursors' masses by our parameter $f_{haze}$ (the mass fraction of precursors that form soots) to find the total mass of the haze particles in a given layer. We vary both $f_{haze}$ and the mode particle size as free parameters, and calculate the optical properties of the haze using Mie theory. 

\subsection{Sulfide/Salt Cloud Model}

To model sulfide and salt clouds, we use a modified version of the \ct{AM01} cloud model \cp{Morley12, Morley13, Morley15}. Cloud material in excess of the saturation vapor pressure of the limiting gas is assumed to condense into spherical, homogeneous cloud particles. We extrapolate the saturation vapor pressure equations from \ct{Morley12} to high metallicites, which introduces some uncertainties but serves as a reasonable first-order approximation for the formation of these cloud species. Cloud particle sizes and vertical distributions are calculated by balancing transport by advection with particle settling.

\subsection{Thermal Emission Spectra }

We use a radiative transfer model developed in \ct{Morley15} to calculate the thermal emission of a planet with arbitrary composition and clouds. Briefly, this model includes the C version of the open-source radiative transfer code \texttt{disort} \cp{Stamnes88, Buras11} which uses the discrete-ordinate method to calculate intensities and fluxes in multiple-scattering and emitting layered media. 

\subsection{Albedo Spectra}

We calculate albedo spectra following the methods described in \ct{Toon77, Toon89, Mckay89, Marley99, Marley99b, Cahoy10}. Here, we use the term geometric albedo to refer to the albedo spectrum at full phase ($\alpha$=0, where the phase angle $\alpha$ is the angle between the incident ray from the star to the planet and the line of sight to the observer): 

\begin{equation}
A_g(\lambda)=\dfrac{F_p(\lambda,\alpha=0)}{F_{\odot,L}(\lambda)}
\end{equation}    
where $\lambda$ is the wavelength, $F_p(\lambda,\alpha=0)$ is the reflected flux at full phase, and $F_{\odot,L}(\lambda)$ is the flux from a perfect Lambert disk of the same radius under the same incident flux.

 \begin{figure}[tb]
  \center   \includegraphics[width=3.7in]{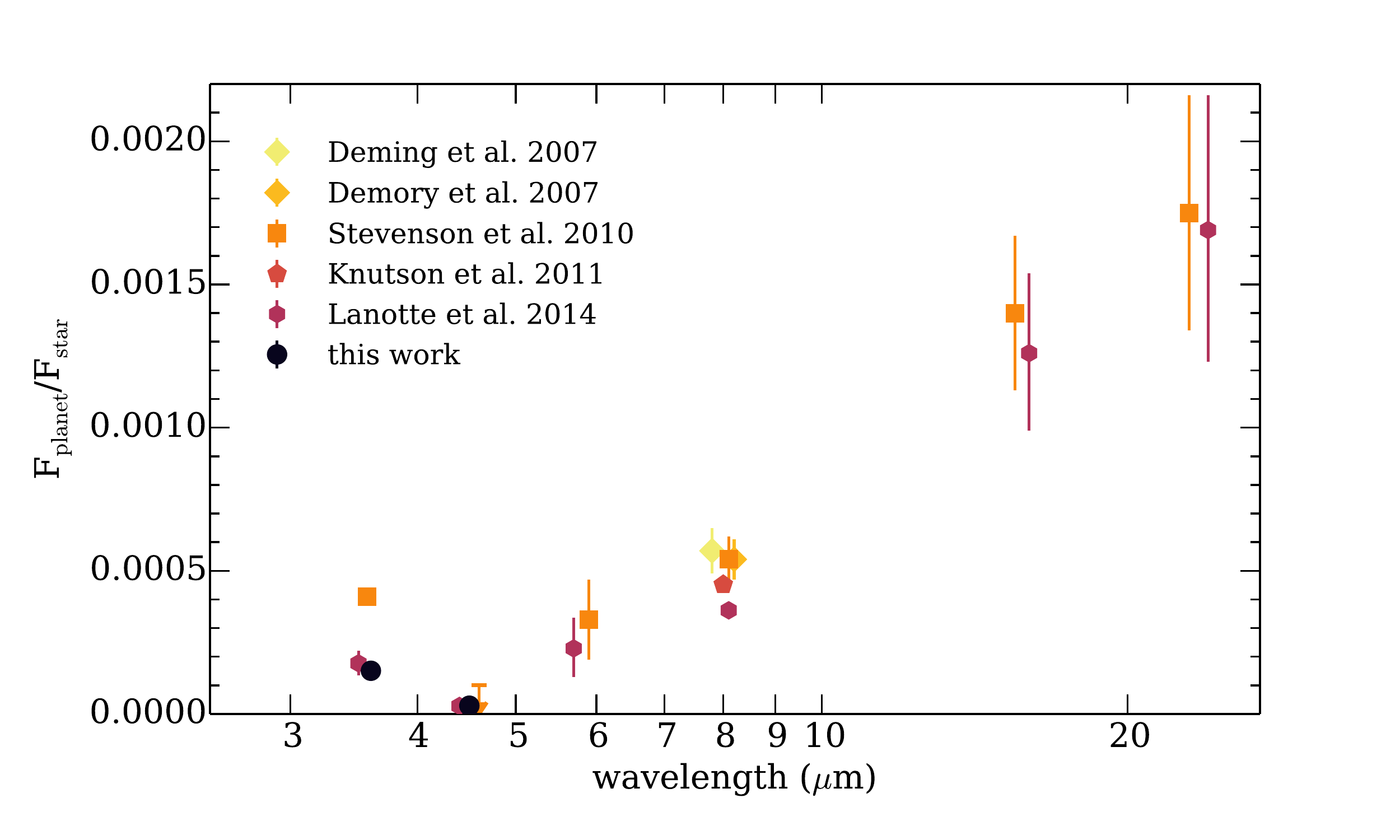}
 \caption{Eclipse depths in the 6 \emph{Spitzer} bandpasses from the literature and this work. Different publications are offset slightly in wavelength for clarity; darker colors indicate later years. }
\label{dataplot_therm}
\end{figure}

\subsection{Retrieval Model}

To more thoroughly explore the chemically plausible parameter space allowed by the emission spectrum, we employ the chemically consistent atmospheric retrieval scheme described in \ct{Kreidberg15} and \ct{Greene16} based on the CHIMERA \cp{Line13a, Line14a} emission forward model. The retrieval uses the 6 parameter analytic radiative equilibrium temperature profile scheme of \ct{Parmentier13} (see \ct{Line13a} for implementation within the emission retrieval) where the free parameters are the infrared opacity ($\kappa_{IR}$), the ratio of the visible to infrared opacity for two visible streams ($\gamma_1$, $\gamma_2$), the partitioning between the two visible streams ($\alpha$), scaling to the top-of-atmosphere irradiation temperature ($\beta$, to accommodate for the unknown albedo and redistribution), and finally the internal temperature (T$_{\rm int}$). These parameters are all free parameters, not recalculated to be consistent with the derived abundances \cp{Line12}. 

The molecular abundances are initially computed along the temperature profile under the assumption of thermochemical equilibrium (using the Chemical Equilibrium with Applications routine, Gordon \& McBride 1994;1996; \ct{Line10, Moses11, Line11}) given the bulk atmospheric metallicity ([M/H]) and carbon-to-oxygen ratio (C/O).  To account for possible disequilibrium chemistry we include a "quench pressure" parameter (P$_{\rm quench}$) whereby the abundances of H$_2$O, CH$_4$, and CO above the quench are fixed at their quench pressure values, a valid representation of many disequilibrium models \cp[e.g.,][]{Moses11, Line11, Zahnle14}. The temperature profile and chemistry parameters result in a total of 9 free parameters. Bayesian estimation is performed using a multi-modal nested sampling algorithm \cp{Feroz09} implemented with the PYMULTINEST routine \cp{Buchner14} recently employed in \ct{Line16}, with generous uniform priors on each parameter (see Table \ref{priortable}).

\begin{deluxetable}{ll}[b!]
\tabletypesize{\scriptsize}
\tablecolumns{2} 
\tablewidth{0pt}
\tablecaption{Uniform prior ranges on the retrieved parameters\label{priortable}}
\tablehead{
\colhead{parameter} & \colhead{range}
}  
\startdata
log($\kappa_{IR}$)[cm$^2$/g]           &      $-3$ to 0   	\\
log($\gamma_1$, $\gamma_2$)       &      $-3$ to 2	\\	
$\alpha$                           &     0 to 1		\\
$\beta$                           &      0 to 2	\\
T$_{\rm int}$ (K)           &     100 to 400	\\
M/H 	&     $10^{-4}$ to $10^4 \times$ solar  \\
log(C/O)$^{\rm a}$                               &      $-2$ to 2  \\
log(P$_{\rm quench}$) [bar]        &       $-6$ to 1.5 	\\
\enddata
\tablenotetext{a}{Solar log(C/O) is $-$0.26.}

\end{deluxetable}

\begin{figure}[tb]
 \includegraphics[width=3.6in]{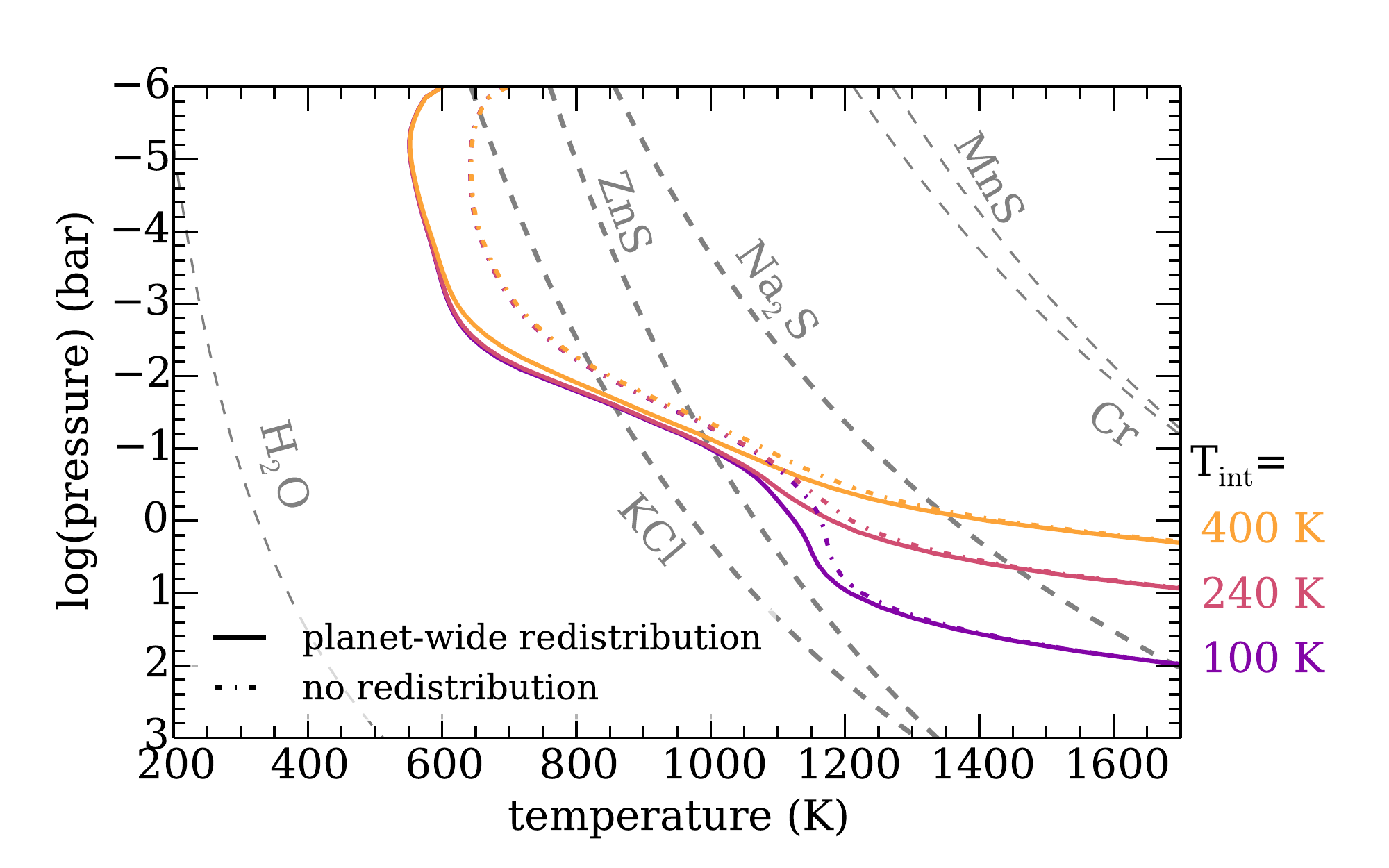} 
 \caption{Pressure--temperature profiles with condensation curves. All models are cloud-free with 300$\times$ solar composition. Solid lines show models with \tint=100, 240, and 400 K and planet-wide heat redistribution. Dash-dot lines show models with the same \tint s but with no heat redistribution (dayside temperature). Condensation curves show where the vapor pressure of a gas is equal to the saturation vapor pressure; cloud material condenses where the P--T profile intersections a condensation curve. }
\label{ptprof_clouds}
\end{figure}

\begin{figure*}[t]
\center \includegraphics[width=4.7in]{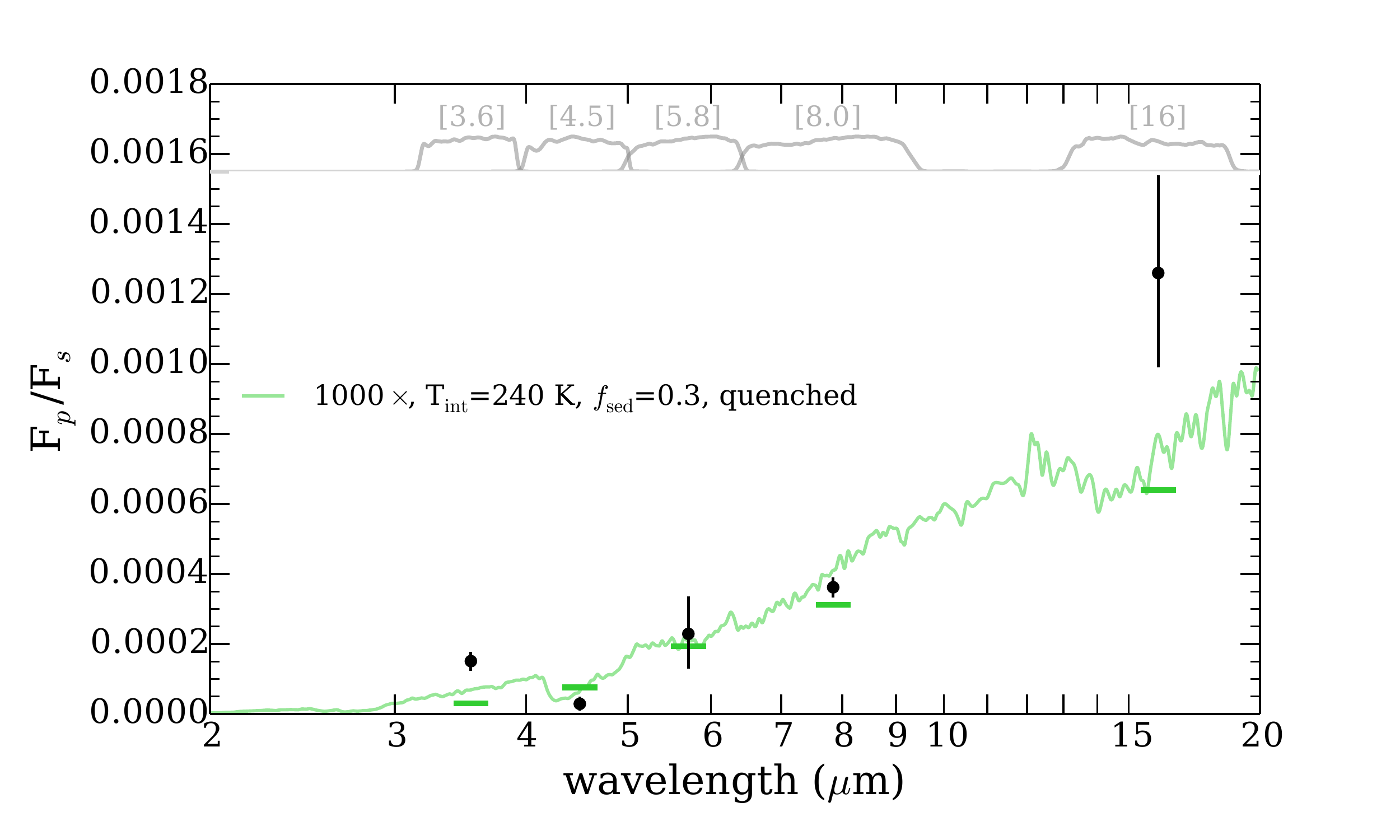}
\vspace{-0.25in}
\center \includegraphics[width=4.7in]{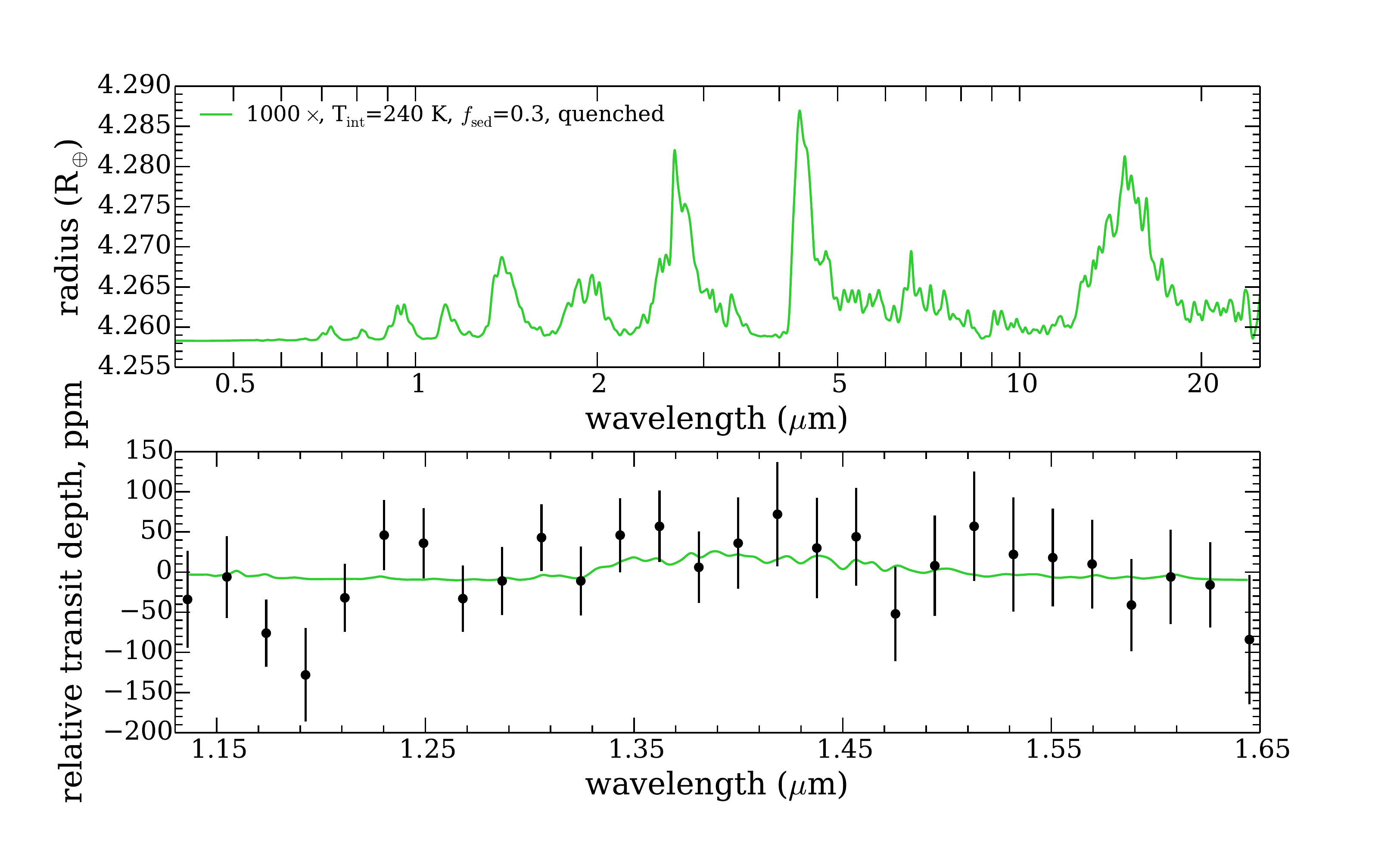}
 \caption{Best-fit thermal emission and transmission spectra. Top panel: Thermal emission spectrum of the best-fit model from the suite of forward models compared to the data. The model is shown as a green line, with synthetic model photometry shown as horizontal lines at the central wavelength of the filter. Data are shown as black points with 1-$\sigma$ error bars. The filter functions for the photometry are shown as gray lines in the top panel. Bottom panels: Transmission spectrum of the same best-fit thermal emission model from the suite of forward models compared to the data. The model is shown as a green line in both panels. The HST/WFC3 transmission spectrum is shown as black points with 1-$\sigma$ error bars in the bottom panel.   }
\label{bestmodel_therm}
\end{figure*}

\section{Results}

\subsection{Observations}

The new eclipse depths are shown in Table \ref{eclipse_param} and Figure \ref{dataplot_therm}. Our eclipse depths of 151$\pm$27 ppm at 3.6 \micron\ and 29$^{+20}_{-16}$ ppm at 4.5 \micron\ are consistent to 1-$\sigma$ with those published in \ct{Lanotte14} (177$\pm$45 and 28$^{+25}_{-18}$ ppm respectively), with a moderate reduction in the uncertainties in both bands. This result serves as confirmation of the high flux at 3.6 \micron\ compared to 4.5 \micron.

\subsection{Self-Consistent Modeling} \label{selfconsistent}

We ran a variety of models from 50--1000$\times$ solar metallicity, varied heat redistribution (planet-wide average and dayside average), internal temperatures (T$_{\rm int}$) from 100--400 K, with clouds (\fsed=0.01--1), and hazes with particle sizes from 0.01--1 \micron\ and \fhaze\ from 1--30\%. We compare each model to the thermal emission photometry from this work (3.6 and 4.5 \micron) and from \ct{Lanotte14} (5.6, 8.0, 16 \micron), using a chi-squared analysis to assess relative goodness-of-fit between the models.  

We show example pressure--temperature profiles along with cloud condensation curves in Figure \ref{ptprof_clouds}. Raising the internal temperature, \tint, increases the temperature of the deep atmosphere (P$\gtrsim$0.1 bar). The heat redistribution of incident stellar flux controls the temperature in the upper atmosphere. GJ 436b's profile crosses condensation curves of sulfides and salts, suggesting that if the atmosphere is cloudy, those clouds may be composed of \nas, KCl, and ZnS.

\subsubsection{Best-fit fiducial model}

Of the 288 models in our grid of cloudy and cloud-free planets, our nominal best-fit set of parameters are: 
\begin{itemize}
	\item 1000$\times$ solar metallicity 	
	\item T$_{\rm int}$=240 K 			
	\item \fsed=0.3 sulfide/salt clouds	
	\item disequilibrium chemistry via quenching		
	\item full heat redistribution (planet-wide average PT profile)
\end{itemize}

This model provides an excellent fit to the transmission spectrum ($\chi^2_{\rm red}<$1 assuming 3 degrees of freedom), though an inadequate fit to the thermal emission ($\chi^2_{\rm red}\sim$11 assuming 3 degrees of freedom). We show the thermal emission and transmission spectra in Figure \ref{bestmodel_therm}.

\subsubsection{Equilibrium and disequilibrium chemistry}

As has been discussed in the literature \cp{Stevenson10, Line11, Moses13}, GJ 436b's high 3.6 \micron\ flux and low 4.5 \micron\ flux indicate that it likely has a high abundance of CO and CO$_2$ relative to CH$_4$. Since equilibrium chemistry for an object at  GJ 436b's temperature would instead result in high abundances of CH$_4$ at metallicities similar to Neptune, this indicates that GJ 436b's chemistry is in disequilibrium. This disequilibrium may be due to a combination of vertical mixing, photochemistry, and other effects \cp{Line11}. Here, we approximate the effect of disequilibrium chemistry by `quenching' the abundances of the carbon species (CO, CO$_2$, CH$_4$) in the atmosphere at deep pressures (10 bar), effectively setting the abundances of these species to be constant through the atmosphere. 

\begin{figure}[tb]
\center \includegraphics[width=3.7in]{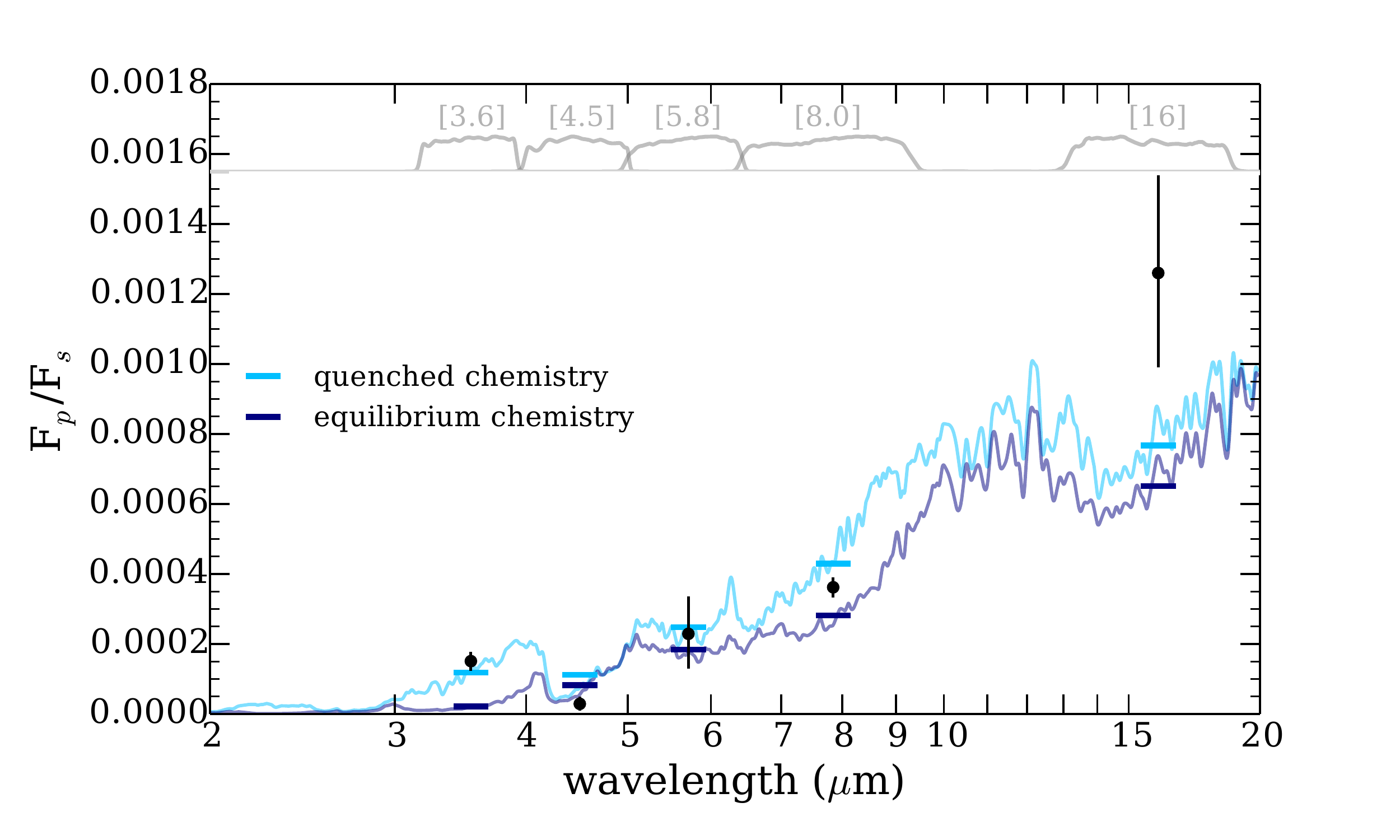}
 \caption{Effect of chemistry on thermal emission spectrum. Both models assume 300$\times$ solar metallicity, \fsed=1 sulfide/salt clouds, planet-wide heat redistribution, and \tint=240 K. The darker blue line and horizontal bars show a model spectrum and photometry assuming equilibrium chemistry; the lighter blue line and horizontal bars show the same model, but with the chemistry quenched at the 10 bar abundances throughout the atmosphere. }
\label{spectra_thermal_quench}
\end{figure}

The resulting effect of disequilibrium chemistry on spectra is shown in Figure \ref{spectra_thermal_quench}. In equilibrium, the model predicts that GJ 436b would be very faint at 3.6 \micron, and progressively brighter at redder wavelengths. In disequilibrium, as is observed in the data, the planet is predicted to be brighter at 3.6 \micron\ due to decreased absorption by CH$_4$. In general, even the models that include disequilibrium chemistry overpredict the brightness at 4.5 \micron\ compared to the observed flux, despite the higher abundance of CO and CO$_2$ in disequilibrium.

\subsubsection{Metallicity}

\begin{figure}[tbh]
\center \includegraphics[width=3.7in]{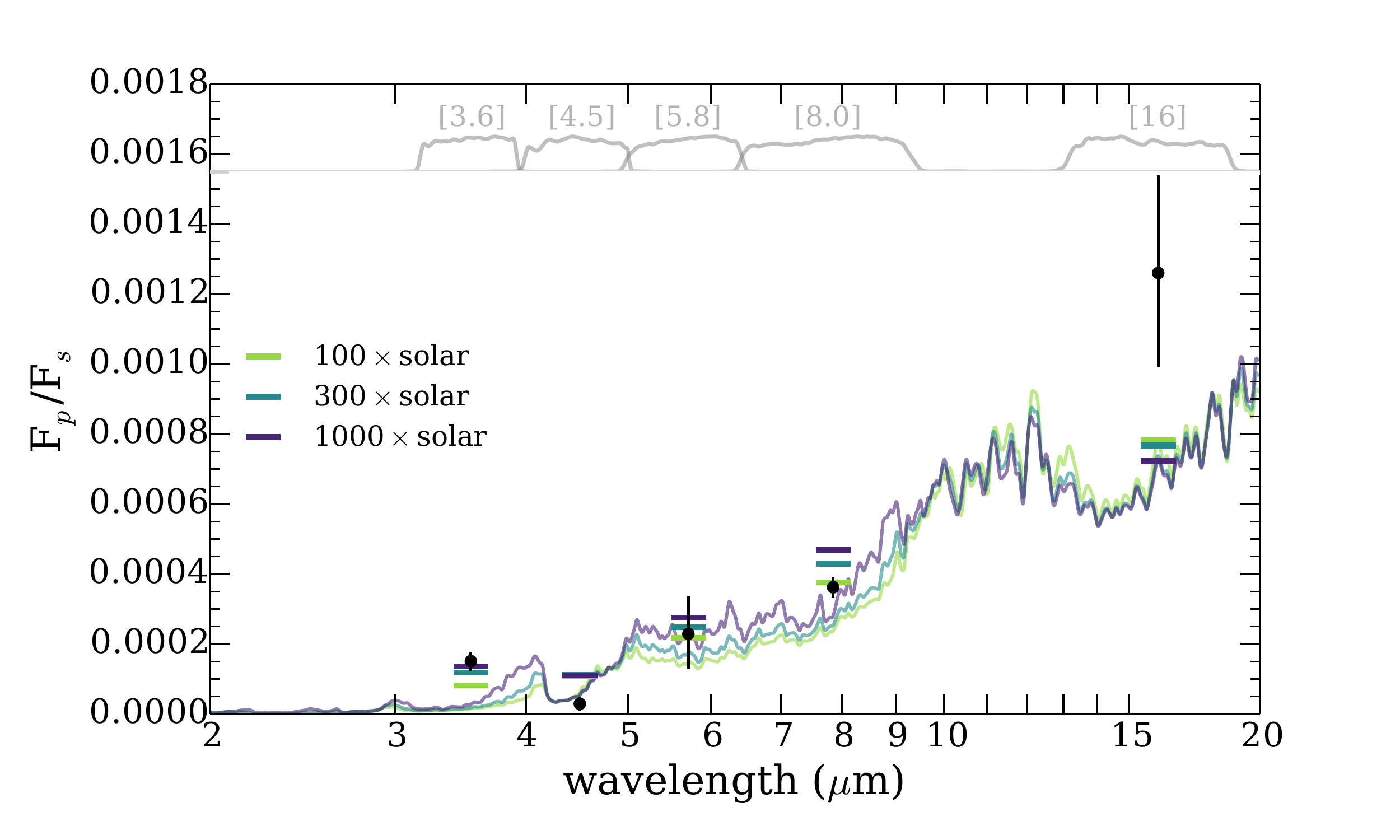}
 \caption{Effect of metallicity on thermal emission. Each model assumes equilibrium chemistry, \fsed=1 sulfide/salt clouds, planet-wide heat redistribution, and \tint=240 K. Metallicities of 100, 300, and 1000$\times$ solar metallicity are shown. Increasing the metallicity decreases the CH$_4$ abundance and increases CO and CO$_2$ abundance. }
\label{spectra_thermal_met}
\end{figure}

Increasing the metallicity of GJ 436b's atmosphere allows us to fit both the thermal emission and transmission spectrum more accurately. There are two reasons for this. As has been discussed at length in \ct{Moses13}, high metallicity atmospheres are predicted, in equilibrium or disequilibrium, to have higher abundances of CO and CO$_2$ relative to CH$_4$. Pushing the chemistry to CO/CO$_2$-rich compositions is crucial to match GJ 436b's thermal emission. We show this effect in Figure \ref{spectra_thermal_met}; models at high metallicities have higher flux at 3.6 and 8 \micron\ due to the change in chemistry. We find that this effect partially saturates at metallicities greater than 300$\times$ solar. 

\begin{figure}[tbh]
\center \includegraphics[width=3.6in]{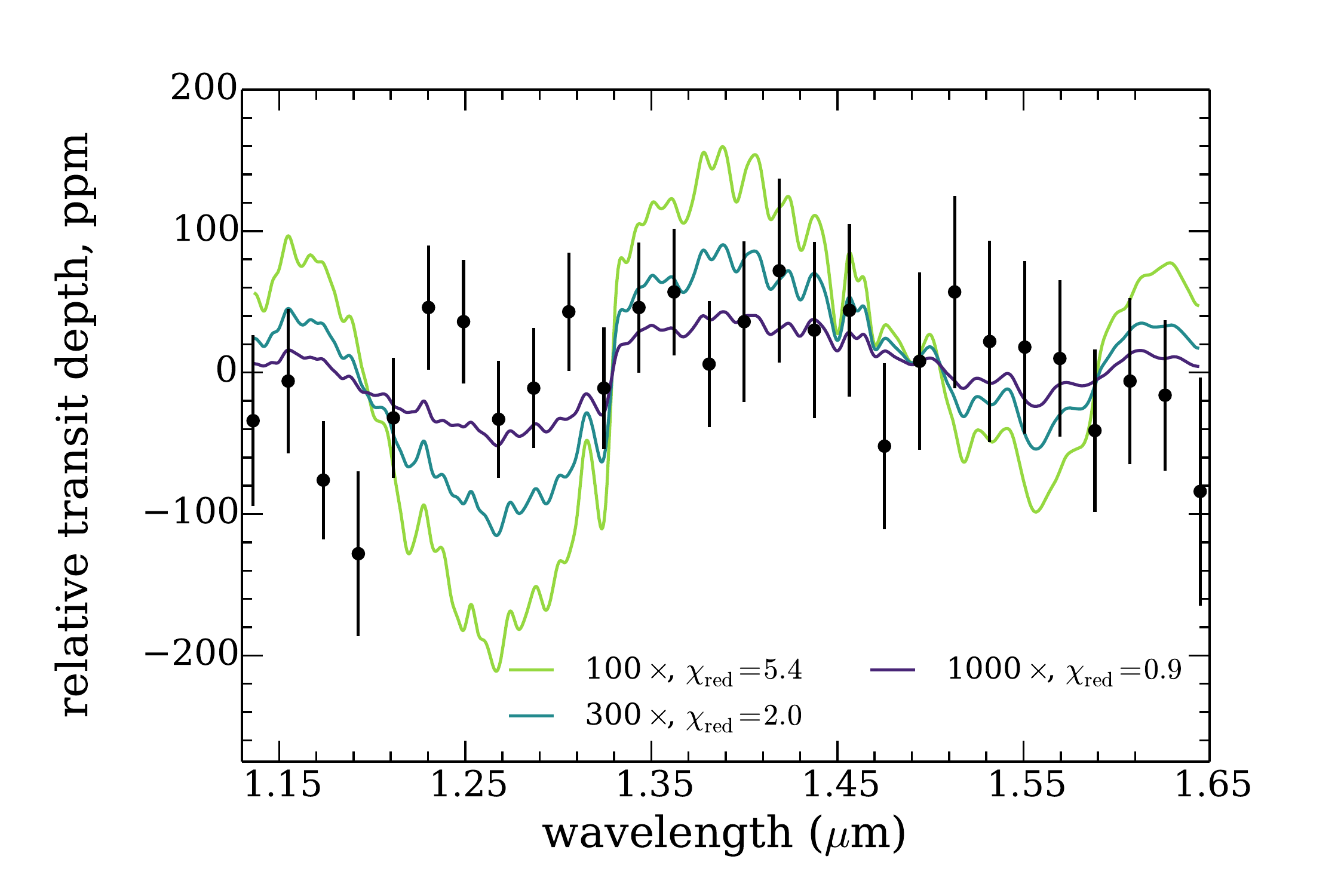}
 \caption{Effect of metallicity on transmission spectrum. Each model is cloud-free, with planet-wide heat redistribution, equilibrium chemistry, and \tint=240 K. Metallicities of 100, 200, 300, and 1000$\times$ solar metallicity are shown. Increasing the metallicity decreases the CH$_4$ abundance and increases CO and CO$_2$ abundance. }
\label{spectra_trans_met}
\end{figure}

High metallicities also make it much easier to flatten the transmission spectrum of GJ 436b sufficiently to match the featureless HST/WFC3 transmission spectrum even in the absence of clouds \cp{Knutson14a}. In Figure \ref{spectra_trans_met} we show cloud-free models for different metallicities. While at metallicities lower than 1000$\times$ solar metallicity clouds are required to sufficiently flatten the spectrum, for models above 1000$\times$ solar metallicity even cloud-free models have high enough mean molecular weights that the size of the features, which scale according to the scale height of the atmosphere, are small enough that they appear featureless at the S/N of the data. 

\subsubsection{Tidal heating}

As a Neptune-sized planet orbiting an old star, without an additional energy source, GJ 436b's interior temperature \tint\ would be $\sim$60 K, slightly warmer than Neptune which has a \tint$\sim$50 K \cp{Fortney07a}. However, GJ 436b is on an eccentric orbit (e$\sim$0.15) despite orbiting its star at a semi major axis where it is predicted to have a tidally circularized orbit, indicating that its interior may still be heated by tidal dissipation. \ct{Moses13} and \ct{Agundez14} both considered the effect of tidal heating, noting that a hotter interior changes the chemistry of the deep interior and therefore the resulting emission spectrum. 

Increasing \tint\ tends to move the deep P--T profile (see Figure \ref{ptprof_clouds}) to regions with high CO/CO$_2$ and lower CH$_4$ abundances, which allows us to better match the observed spectrum. Heating the deep atmosphere also increases the effective temperature of the atmosphere by changing the P--T profile, increasing flux at all \emph{Spitzer} wavelengths. This effect is shown in Figure \ref{spectra_thermal_tint} for three different \tint\ values (100, 240, and 400 K). Best-fit models cluster around \tint=240 K, a temperature that allows us to match the 3.6, 5.6, and 8.0 \micron\ points relatively well, while over predicting the 4.5 \micron\ flux somewhat. 

We note that this is the first indication that the internal temperature of a planet has an important and observable effect on the emission spectrum of a transiting planet. 

\begin{figure}[tbh]
\center \includegraphics[width=3.6in]{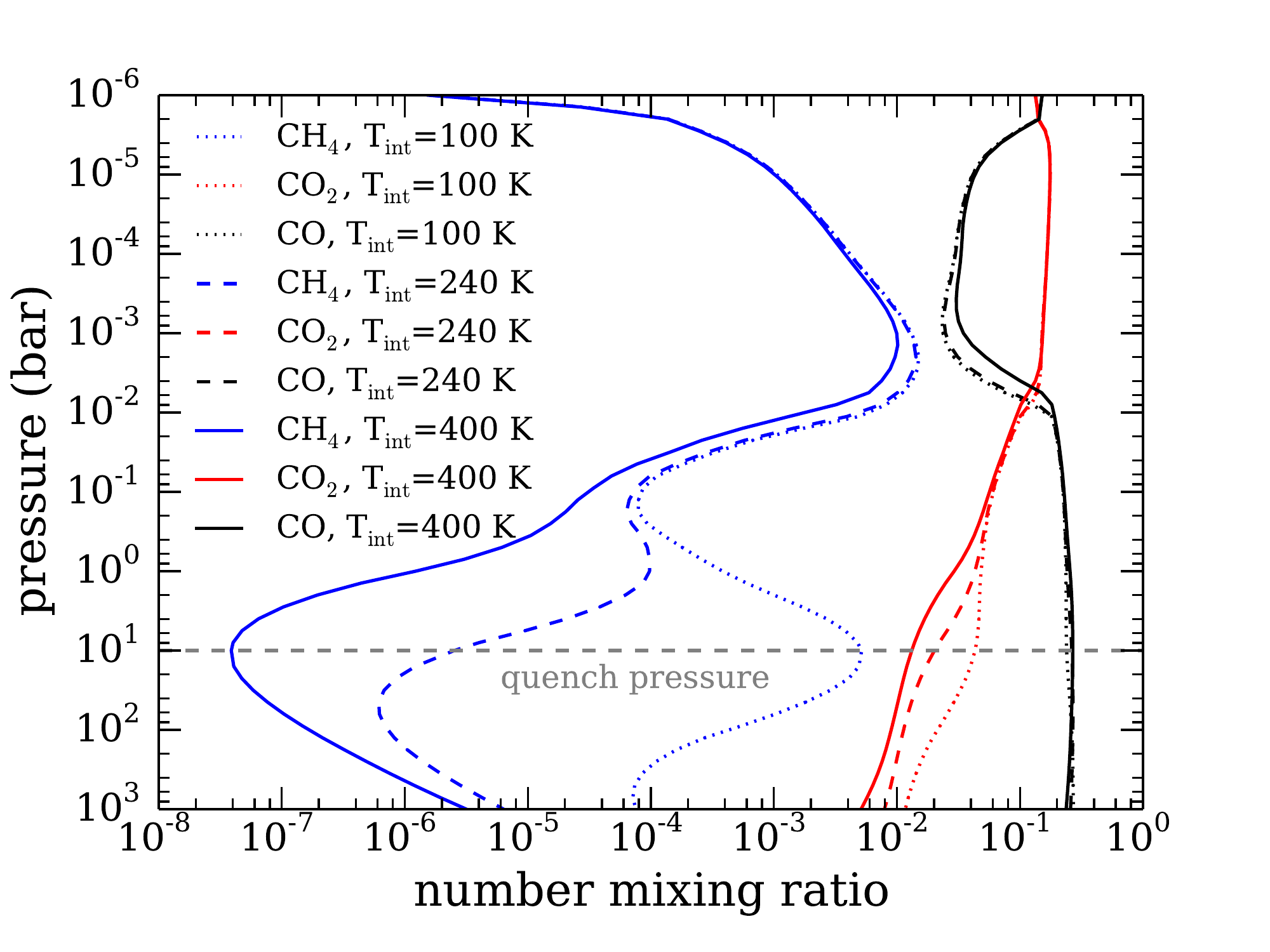}
 \caption{Abundances of major carbon-bearing species in chemical equilibrium. All models have a composition of 1000$\times$ solar metallicity and a planet-wide average PT profile. Different \tint\ values are shown with different line styles, and each molecule (CH$_4$, CO, CO$_2$) is shown in a different color. The fiducial quench pressure used in the self-consistent modeling is shown as a horizontal dashed line. Note that increasing the internal temperature decreases the CH$_4$ abundance in the deep atmosphere. }
\label{chemistry}
\end{figure}

\begin{figure}[tb]
\center \includegraphics[width=3.7in]{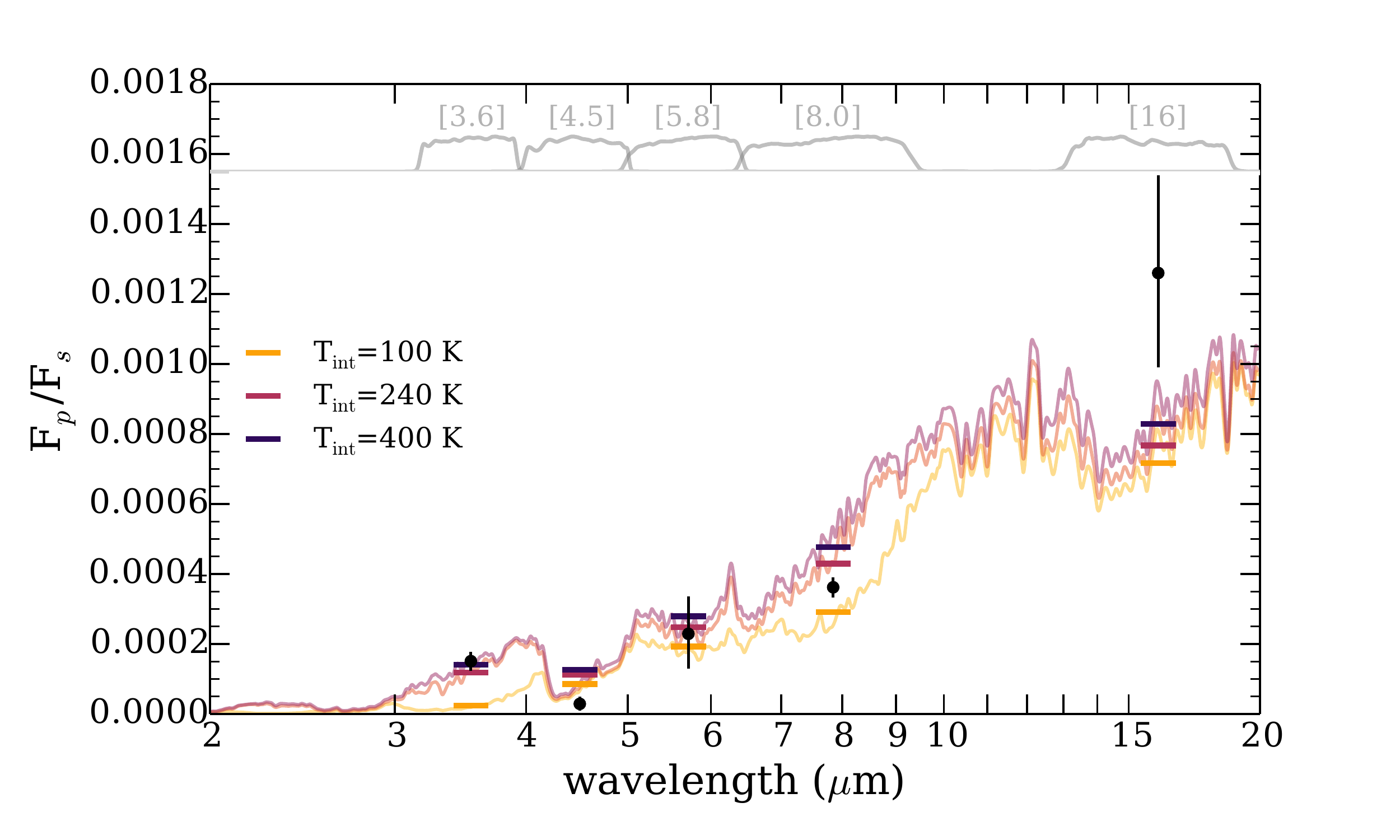}
 \caption{Effect of tidal heating on thermal emission. Each model assumes 300$\times$ solar metallicity, quenched chemistry, \fsed=1 sulfide/salt clouds, and planet-wide heat redistribution. The tidally heated atmospheres (240 and 400 K) have higher abundances of CO and CO$_2$ and lower abundances of CH$_4$ due to the hotter deep atmosphere (where the chemistry is quenched). Tidal heating also increases the \teff\ of the planet by changing the temperature profile, increasing the emergent flux at all wavelengths. }
\label{spectra_thermal_tint}
\end{figure}

\subsubsection{Clouds}

Clouds increase opacity across all wavelengths as (relatively) gray absorbers. This means that including clouds decreases flux between absorption features (e.g. at 3.6 and 8.0 \micron\ for GJ 436b's composition) and somewhat less significantly at the locations of absorption features where the planet is already dark. Thinner clouds (\fsed=0.3--1 in our parameterization) alter the spectrum slightly, while thicker clouds (\fsed$\le$0.1) create a blackbody-like spectrum with the temperature of the top of the cloud. Comparing this to the observed photometry of GJ 436b, these thick clouds significantly under predict the flux at 3.6 \micron\ especially.

\begin{figure}[tb]
\center \includegraphics[width=3.7in]{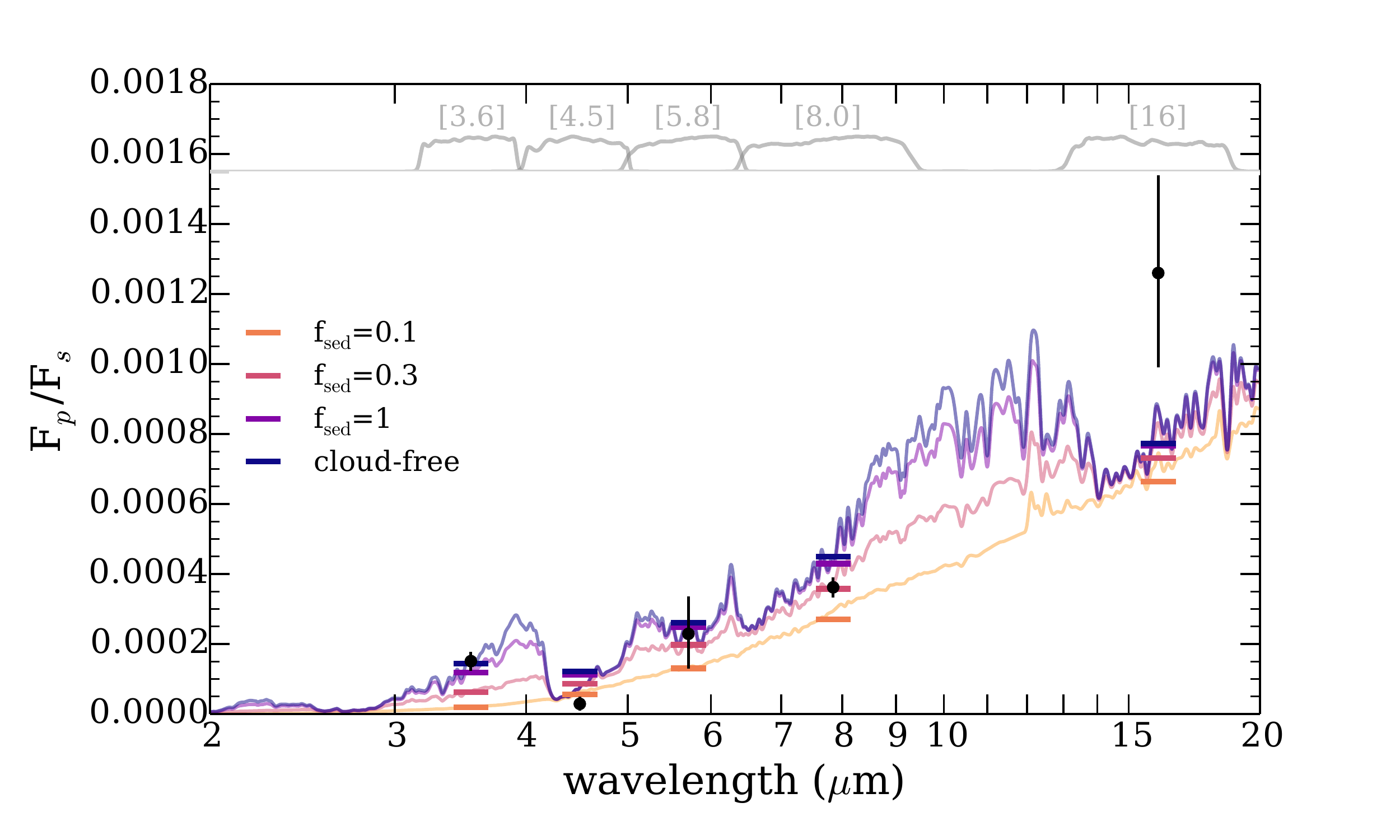}
 \caption{Effect of sulfide/salt clouds on thermal emission. Each model uses the same pressure-temperature profile and assumes 300$\times$ solar metallicity, quenched chemistry, planet-wide heat redistribution, and \tint=240 K. A cloud-free model and cloudy models with \fsed=0.03 to 1 are shown. Cloud opacity decreases the thermal emission across the spectrum. Models with moderate clouds (\fsed=0.3 to 1) fit the \emph{Spitzer} points best. }
\label{spectra_thermal_fsed}
\end{figure}

In transmission, clouds flatten the spectrum without increasing the mean molecular weight of molecular gas in the atmosphere. As discussed above, for metallicities $\sim$1000$\times$ solar, no additional cloud opacity is needed to match the featureless spectrum (\rchi$\sim$1 for all models). At 300$\times$ solar metallicity, thin clouds (\fsed=1) adequately obscure the spectral features, whereas for a Neptune-like 100$\times$ solar composition, \fsed=0.3 clouds are required. In the \ct{AM01} prescription, lower \fsed\ values indicate less efficient sedimentation, causing smaller particles sizes and more lofted clouds. 

\begin{figure}[t]
\center \includegraphics[width=3.6in]{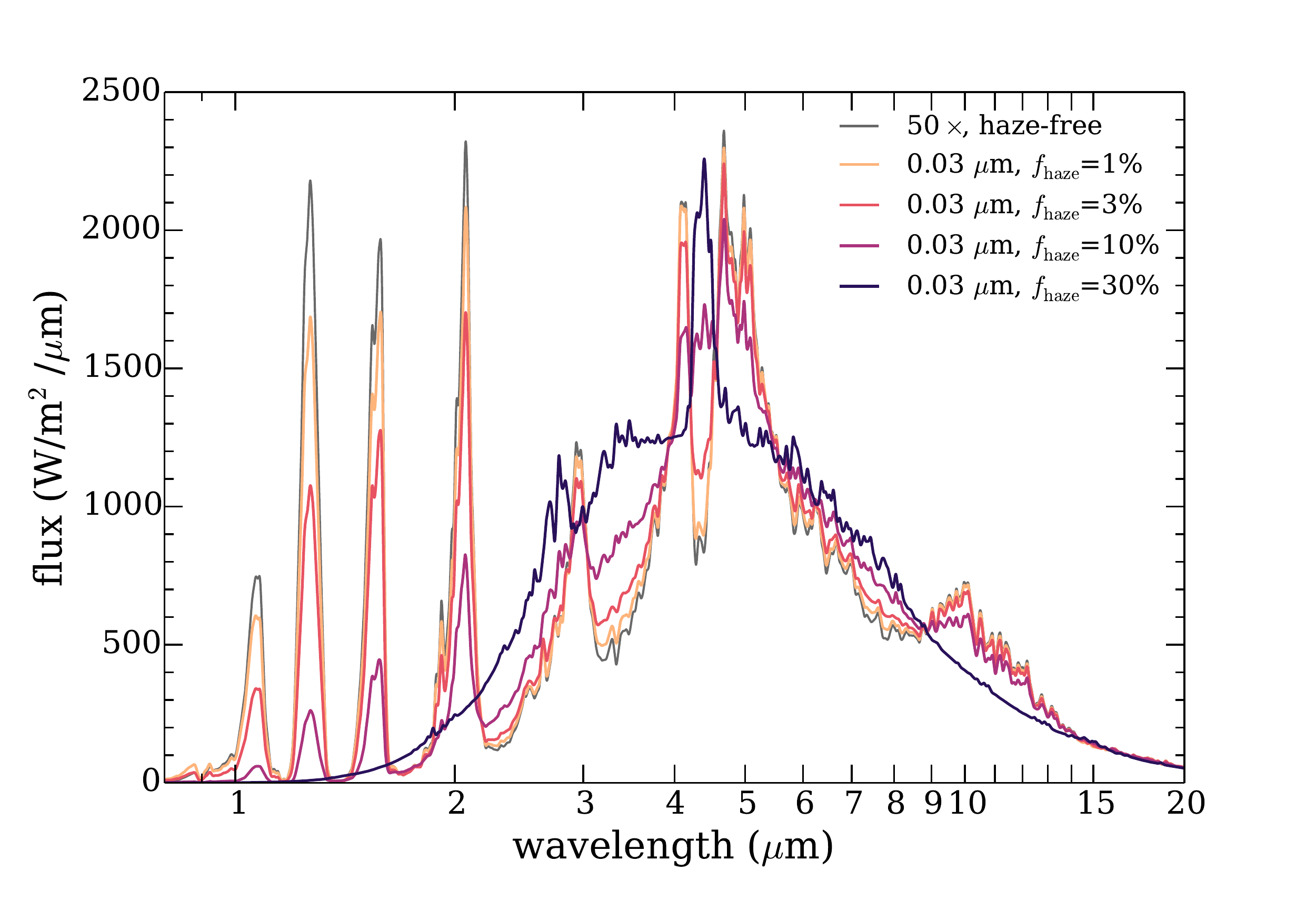}
\vspace{-0.6in}
\center \includegraphics[width=3.6in]{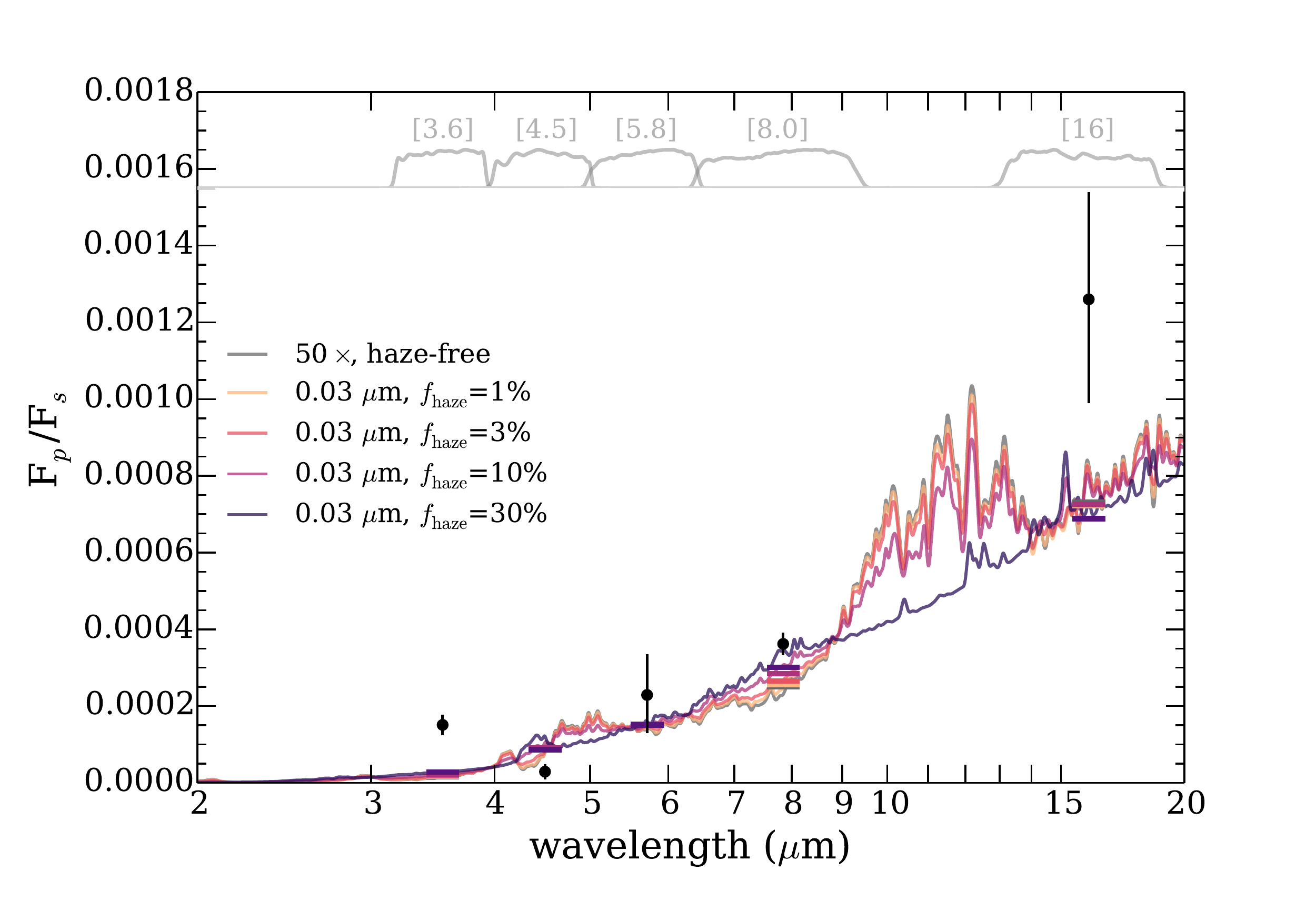}
 \caption{Effect of photochemical hazes on thermal emission. The top panel shows the emergent flux from the planet. All models have 50$\times$ solar metallicity, equilibrium chemistry, and planet-wide heat redistribution. The gray line shows a cloud-free model, and the colored lines show a progression of hazy models with hazy-forming efficiency parameter \fhaze\ varying from 1 to 30\%. The bottom panel shows the same models, but dividing by the flux of the host star to compare to the measured photometry. }
\label{photochemical_haze}
\end{figure}

\begin{figure*}[t]
\center \includegraphics[width=4.5in]{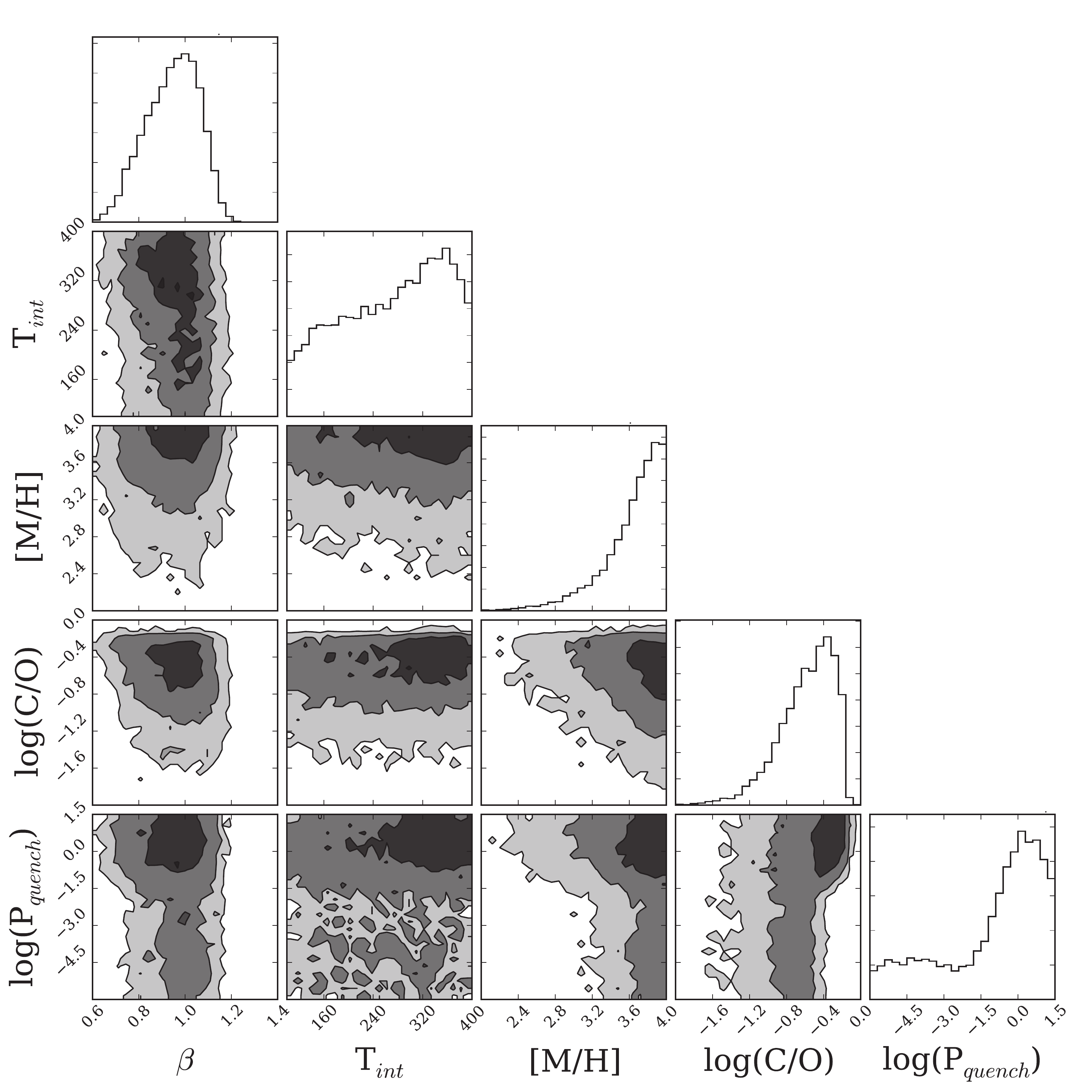}
 \caption{Posterior probability distributions and correlations. The top panel (histogram) shows the posterior probability distribution for each parameter, marginalized over all other parameters. The other panels show 2D contour plots that represent the correlations between each pair of parameters, where the regions from darkest to lightest represent the 1-, 2-, and 3-$\sigma$ contours. }
\label{stairpairs}
\end{figure*}

\subsubsection{Photochemical Hazes in GJ 436b}

We investigate the effect of photochemical hazes on the thermal emission spectrum of GJ 436b. \ct{Morley15} showed that it is possible for optically thick photochemical hazes (such as those postulated to exist in GJ 1214b) to cause a temperature inversion in the upper atmospheres of planets. This can change the spectrum such that molecules that would normally be seen in absorption in a planet without a temperature inversion such as methane are actually seen in emission in an atmosphere with a temperature inversion. We tested whether this process could be happening on GJ 436b and causing the observed thermal emission. 

The results of this investigation are summarized in Figure \ref{photochemical_haze}. The top panel shows the thermal emission of the planet alone. We find that it is possible to create a temperature inversion with dark soot-like photochemical haze in GJ 436b, especially for relatively small particle sizes. As expected, methane is seen in emission, significantly brightening the model spectrum at 3.6 \micron\ compared to a haze-free model. As in \ct{Morley15}, CO$_2$ at 4.3 \micron\ is also predicted to be seen in emission at Neptune-like metallicities (in this case 50$\times$ solar metallicity). In the bottom panel, we show the planet-star flux ratio; here it becomes clear that the hazy model does not fit the observations significantly better than the haze-free model. In particular, the model spectrum is much fainter than the planet's 3.6 \micron\ photometric point. The 4.5 \micron\ flux, despite the significant changes to the shape of the spectrum across the bandpass, remains nearly identical across the range of hazy models tested. 

In general, we find that even though a temperature inversion in a methane-rich atmosphere can increase the 3.6 \micron\ flux, it is not a significant enough effect to match the observed flux, and, furthermore, the flux within the 4.5 \micron\ region can also increase due to emission in the CO$_2$ bandpass. We conclude that photochemical hazes cannot erase the need for an atmosphere with significant CO and CO$_2$ and a low abundance of CH$_4$. This required low-CH$_4$ atmospheric composition, in turn, reduces the likelihood that carbon-based photochemical hazes will be significant in the atmosphere \cp{Fortney13}.

\begin{figure*}[tb]
 \includegraphics[height=2.5in]{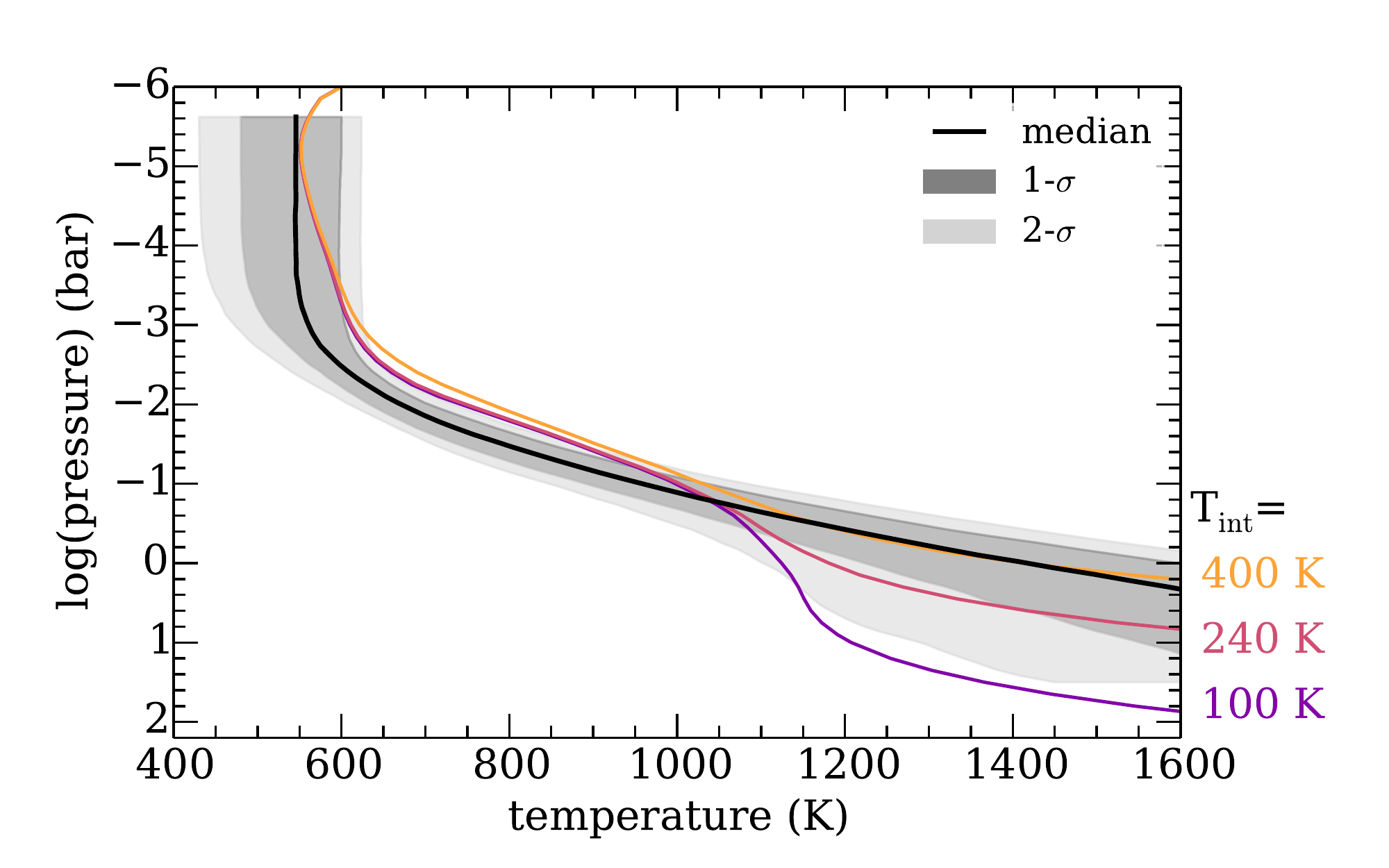}
 \includegraphics[height=2.5in]{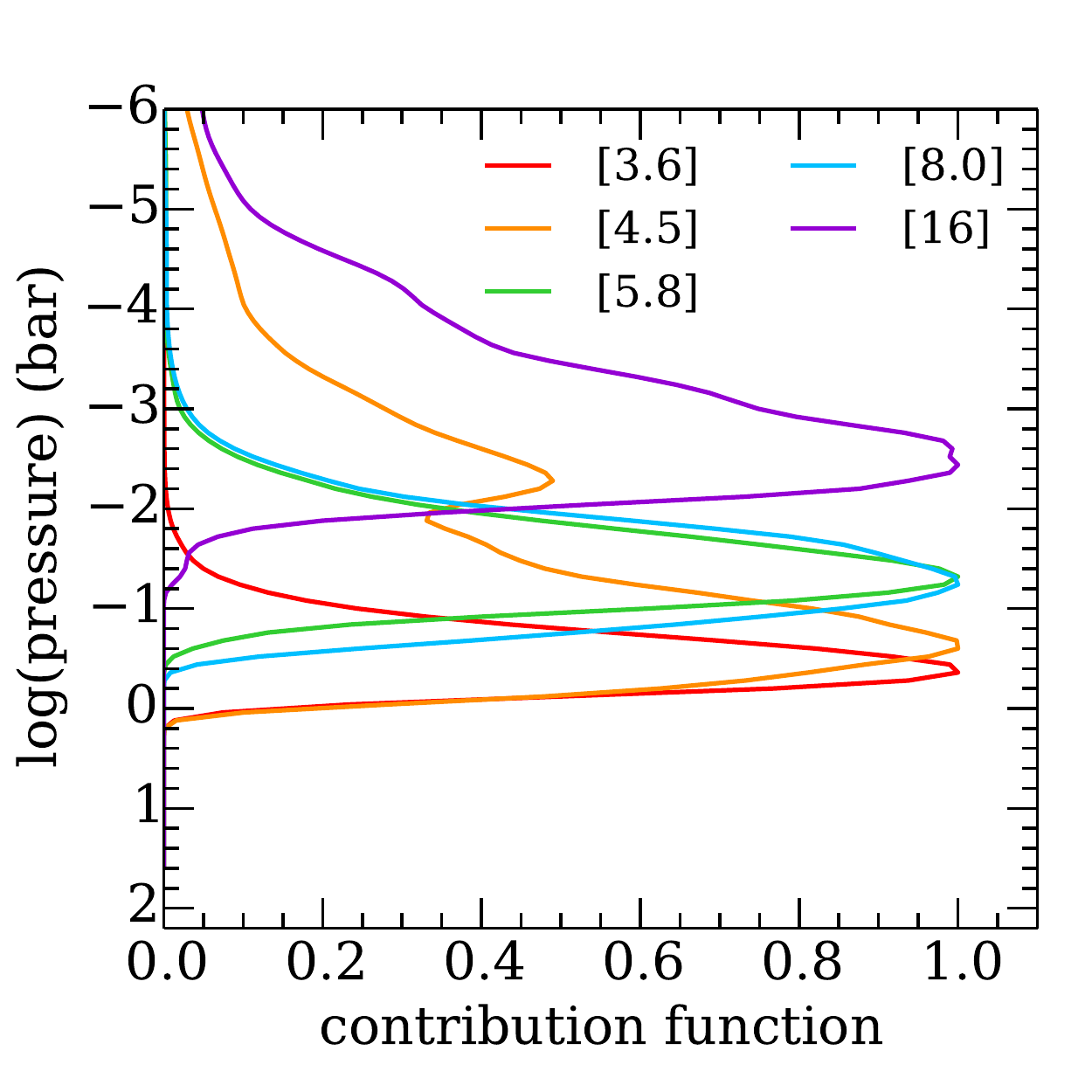}
 \caption{Pressure--temperature profiles and contribution functions for each bandpass. The left panel shows pressure-temperature profiles of both retrieved and self-consistent models. The black line indicates the median retrieved profile while the dark and light gray shaded regions represent the 1- and 2-$\sigma$ confidence regions respectively. The colored lines show self-consistent models with planet-wide heat redistribution and \tint\ of 100, 240, and 400 K. Note the good agreement between the tidally heated (240--400 K) models and the retrieved profile. The right panel shows contribution functions for each of the five bandpasses for a representative retrieval model. The shortest wavelength 3.6 \micron\ band probes the deepest wavelengths while the 16 \micron\ band proves the shallowest. }
\label{pt_retrievals}
\end{figure*}

\begin{figure*}[tbh]
\center \includegraphics[width=5.5in]{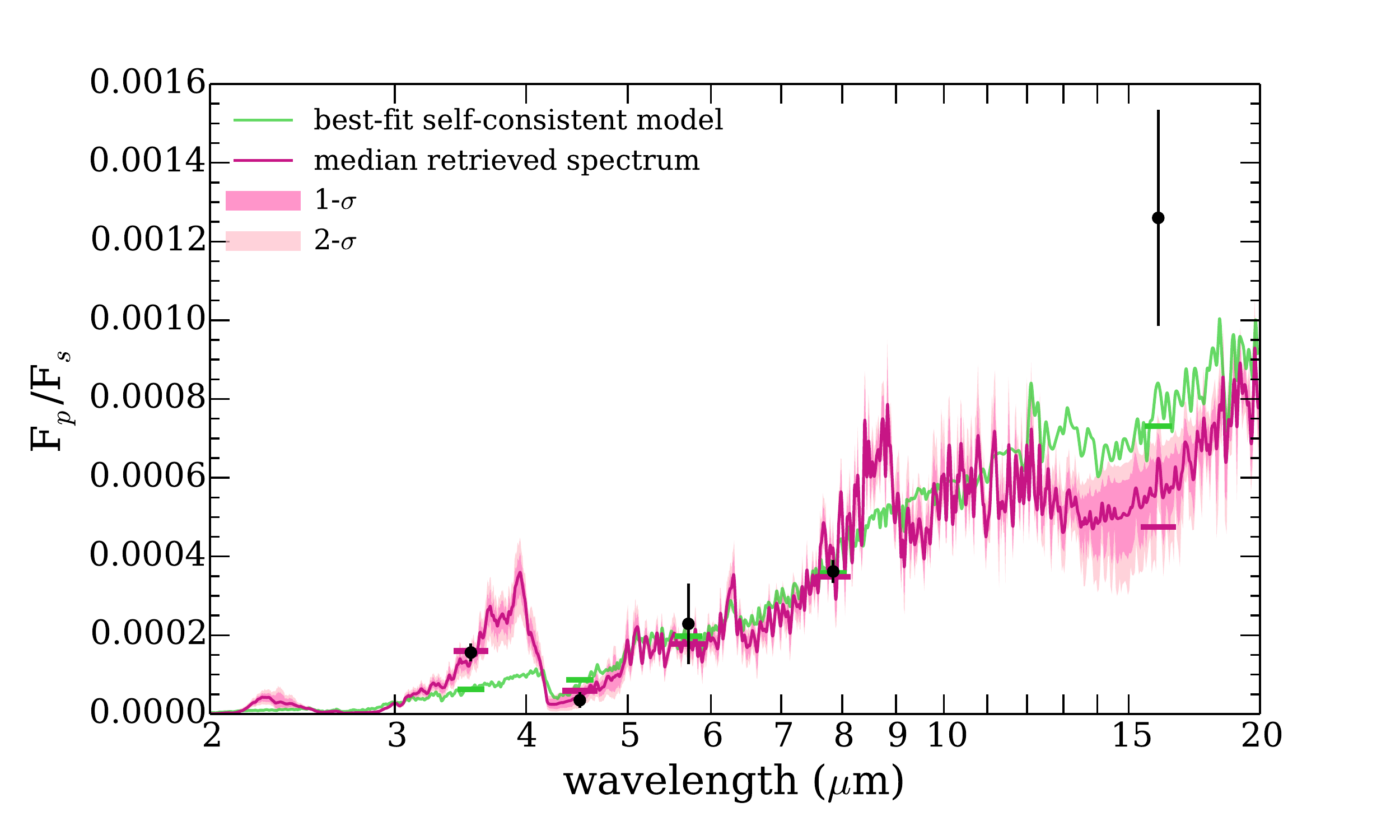}
 \caption{Retrieved data compared to data and best-fit self-consistent model. The pink line and shaded dark and light pink regions are the median fit, 1-$\sigma$, and 2-$\sigma$ confidence intervals respectively. The green line is the best-fit self-consistent model (300$\times$ solar metallicity, \tint=240 K, \fsed=0.3, quenched disequilibrium chemistry).}
\label{spectra_retrievals}
\end{figure*}

\subsection{Retrievals} \label{retrievals}

We have shown in Section \ref{selfconsistent} that we favor models at high metallicity, with both disequilibrium chemistry and tidal heating; these three properties combine to maximize the CO/CO$_2$ abundances and minimize CH$_4$ abundance, allowing the models to match approximately with the measured photometry. Retrieval models provide a quantitative way to test these conclusions and fully explore parameter space beyond our self-consistent model grids. 

We find that retrieval methods draw similar conclusions to the self-consistent modeling; GJ 436b appears to be very high metallicity, with evidence for both deeply-quenched disequilibrium chemistry and thermal heating of the deep interior. For the dayside thermal emission spectrum, the best-fit retrieved solution has a goodness-of-fit divided by number of data points $\chi^2/N$=2.02, compared to $\chi^2/N$=4.54 for the best self-consistent thermal emission spectrum, indicating a significantly improved fit. A comparison of the Bayesian Information Criteria (BIC) reveals that that additional parameters used in the retrieval fitting are necessary, with a BIC of $\sim$26 for the best-fit retrieved solution and $\sim$34 for the best-fit self-consistent solution. 

\subsubsection{Retrieved Posterior Probability Distributions}

Retrieved posterior probability distributions and correlations are shown in the stair-pair plot in Figure \ref{stairpairs} for 5 of the 9 free parameters in the retrieval: $\beta$, \tint, [M/H], log(C/O), log(P$_{\rm quench}$). The best-fit models have:
\begin{itemize}
\item High metallicity. The maximum likelihood model has a metallicity of $\sim$6000$\times$ solar metallicity, with a 3-$\sigma$ lower limit on the metallicity of 106$\times$ solar. 
\item Disequilibrium chemistry. The maximum likelihood model has a quench pressure around 9 bar (with a wide range of values for $P_{\rm quench}$ allowed). 
\item Enhanced internal temperature. The maximum likelihood \tint\ is 336 K (with large uncertainties), indicating that tidal heating may be increasing GJ 436b's internal temperature, in agreement with the tidally heated self-consistent models. 
\item Solar C/O ratio. The maximum likelihood C/O ratio is 0.70, with a sharp cut-off at higher C/O ratios and a long tail to lower C/O ratios.  
\end{itemize}

In Figure \ref{pt_retrievals} we compare the retrieved P--T profile to self-consistent models at 300 $\times$ solar metallicity. We find that retrieved profile is in remarkable agreement with self-consistent models that include the effect of tidal heating in the deep interior. Our best-fit \tint\  from the self-consistent modeling approach (240 K) falls within the 2-$\sigma$ range of the retrieved profile. 

The contribution functions for each of the \emph{Spitzer} bandpasses are also shown in Figure \ref{pt_retrievals}. 3.6 \micron\ probes the deepest pressures, probing pressures as high as 1 bar. As expected, comparing the contribution functions to the range of P--T profiles found by the retrieval, the spread in allowed P--T profiles increases for pressures deeper than 1 bar. The other wavelengths probe lower pressures of the atmosphere, with 5.8 and 8.0 \micron\ centered around 0.05 bar and 16 \micron\ centered around 0.003 bar. The 4.5 \micron\ bandpass has the largest range of pressures, with a peak at deep pressures (0.2 bar) and a long tail to low pressures, unsurprising given that the band covers the spectrum where the modulation is the greatest. 

Figure \ref{spectra_retrievals} shows the best-fit retrieved range of spectra compared to both the data and the best-fit self-consistent model. The retrieved best-fit is statistically and by-eye a somewhat better fit to the data than the self-consistent models. In particular, it has higher flux at 3.6 \micron\ and lower flux at 4.5 \micron. Both the retrieved and self-consistent models fit the 5.6 and 8.0 \micron\ points well; the 16 \micron\ photometry is underestimated by both models, though the error bar is large.

\section{Discussion}

\subsection{Predictions for Reflected Light Spectra}
\begin{figure}[tbh]
\center \includegraphics[width=3.7in]{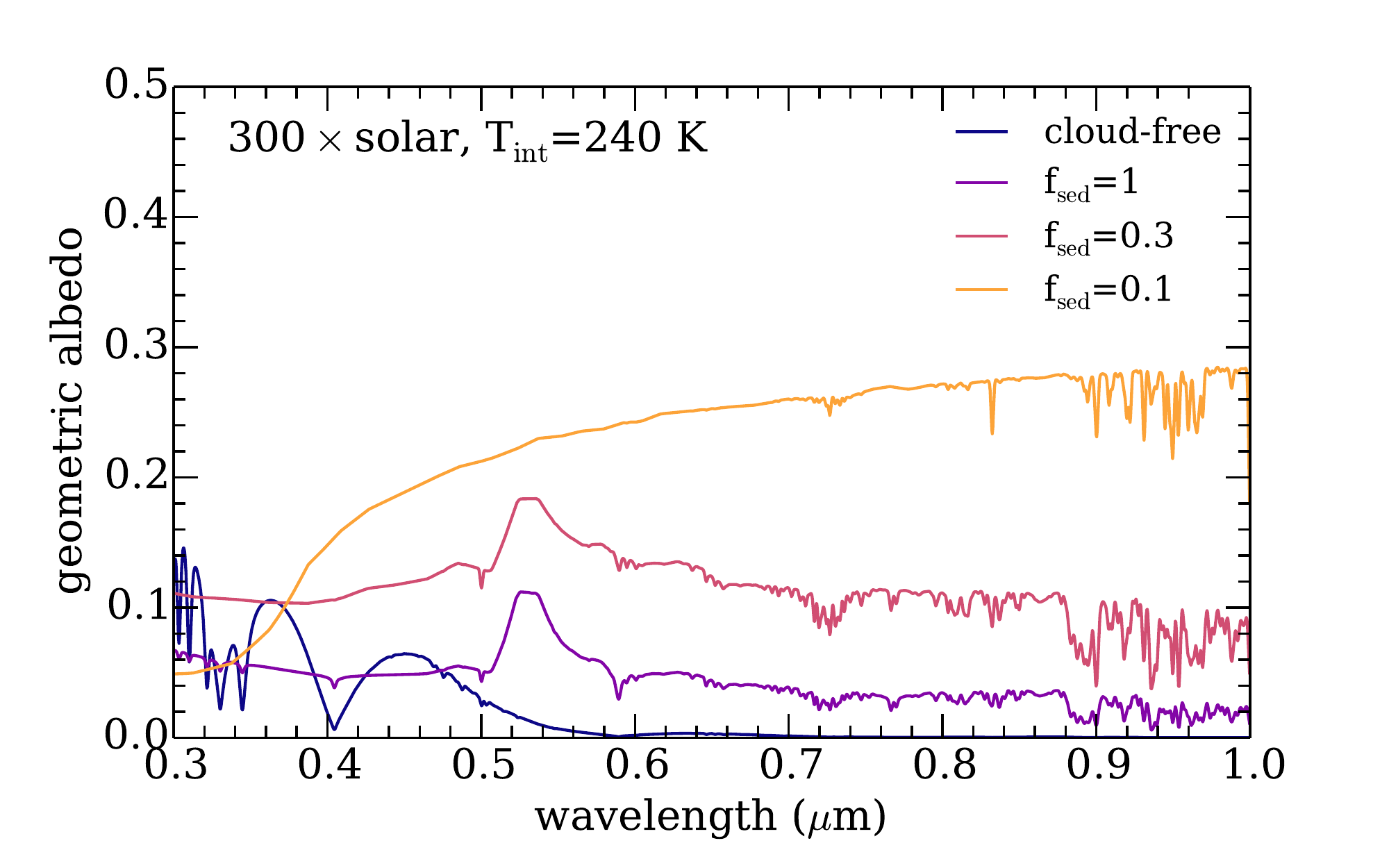}
\vspace{-0.5in}
\center \includegraphics[width=3.7in]{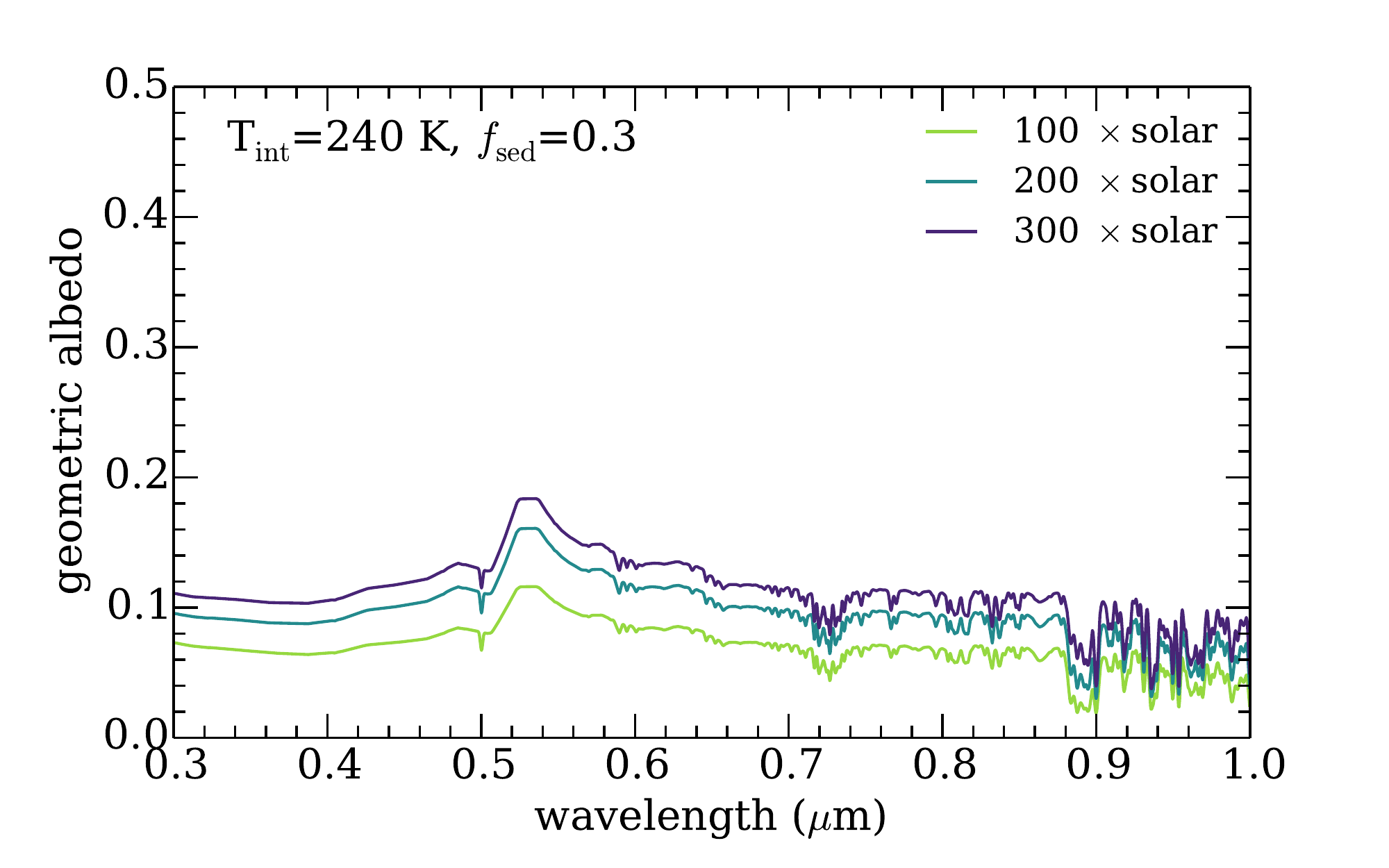}
 \caption{Predicted albedo spectra. Top panel shows models with 300$\times$solar metallicity, \tint=240 K. A cloud-free model and models with cloud parameter \fsed\ from 0.03 to 1 are shown. Bottom panel shows models with \tint=240 K, \fsed=0.3. Metallicities from 100 to 300 $\times$ solar are shown. }
\label{albedo_spectra}
\end{figure}

Cloud properties have the strongest effect on the predicted reflected light spectrum of GJ 436b. Cloud-free models are dark from 0.6--1\micron\ ($A_g$<1\%) and somewhat brighter (up to $A_g\sim$10\%) at bluer wavelengths, as is generally true for cloudless giant planets \cp{Marley99, Sudarsky00}. Thinner clouds (\fsed=0.3--1) are brighter with albedos between a few percent and tens of percent. Thicker clouds (\fsed=0.1) have the brightest albedos from 0.6 to 1\micron, up to nearly 30\%. Some example cloudy spectra are shown in the top panel of Figure \ref{albedo_spectra}. 

Other properties have weaker effects on the reflected light spectrum for this planet. For example, models with metallicities from 100--300$\times$ solar metallicity are shown in the bottom panel of Figure \ref{albedo_spectra}. Increasing the metallicity (which also changes the cloud) increases the geometric albedo across the spectrum.

\subsection{Are Very High Metallicities Reasonable?}

\begin{figure}[tbh]
\center \includegraphics[width=3.5in]{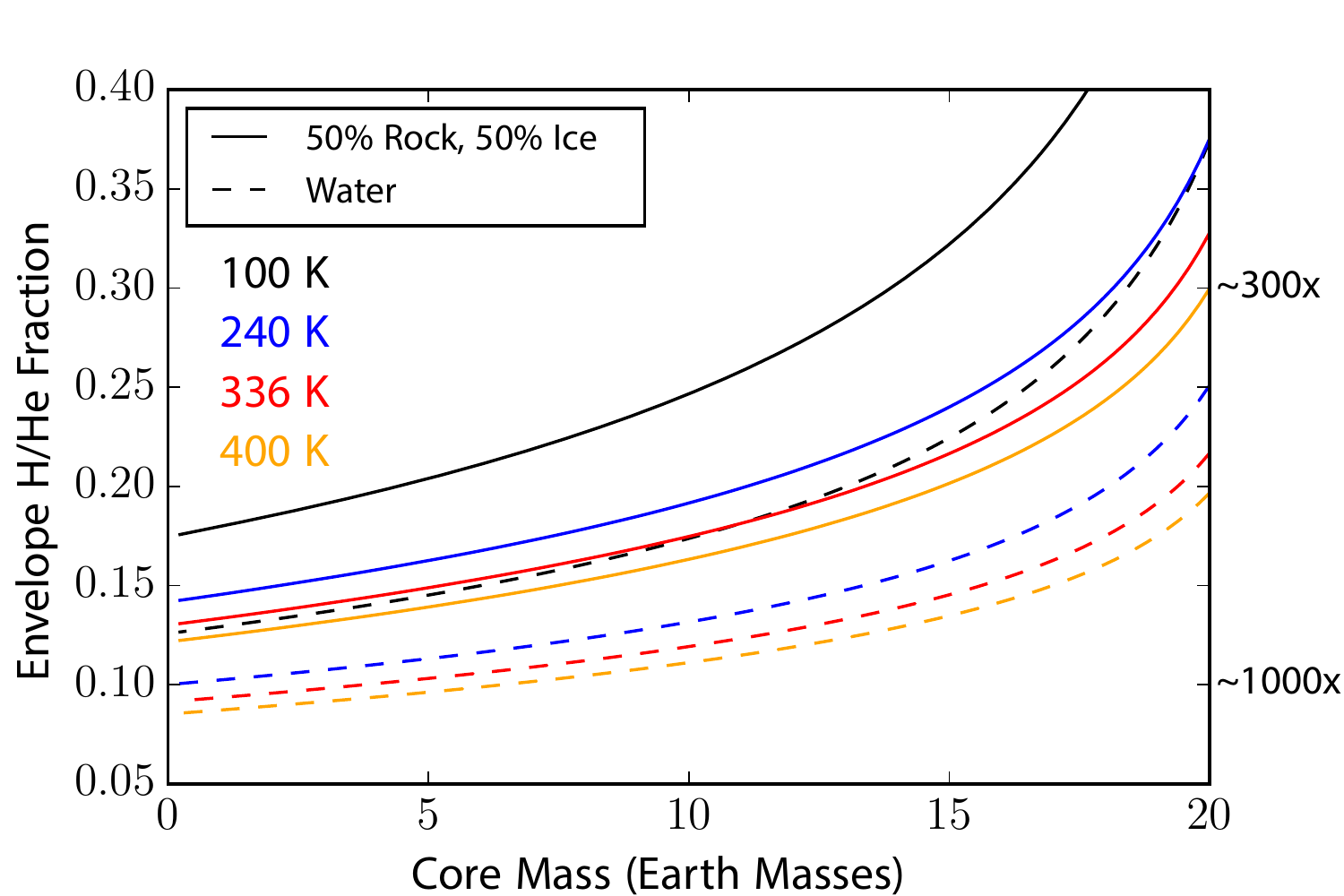}
 \caption{Core mass vs. envelope H/He fraction required to match GJ 436b's measured mass and radius. Models that include a rock/ice core and mix of H/He and rock/ice in the envelope are shown as solid lines, while models that include a water ice core and mix of H/He and water in the envelope are shown as dashed lines. Different colors represent different interior temperature \tint. Approximate conversions between envelope H/He fraction and atmospheric metallicity for 1000 and 300$\times$ solar metallicity are shown at the right. }
\label{interior_models}
\end{figure}

We find that the best-fit atmospheric models have high metallicities, but it remains to be seen whether these values are physically realistic. GJ 436b has a different host star, equilibrium temperature, and orbit than the ice giants in our own solar system, so it likely formed and evolved in very different conditions. The maximum metal-enrichment of the envelope of a Neptune-mass exoplanet is not yet known. Studies of this to date, including \ct{Fortney13}, have suggested that a diverse range of outcomes might be expected for planets in this intermediate mass regime between Earth and Saturn, with potentially high atmospheric enrichments in some cases.

Furthermore, because of the uncertainty in the internal entropy of GJ 436b, its mass and radius do not provide strong limits on the metal-enrichment of the envelope. \ct{Nettelmann10} find that a minimum H/He fraction of 10$^{-3}$ $M_p$ is necessary to match the radius. This very low H/He fraction would require a warm planetary interior, as is favored by the best-fit thermal emission spectra in this work. 

In Figure \ref{interior_models} we show the results from interior models \cp{Thorngren16}, which show the envelope H/He mass fraction required to match GJ 436b's mass and radius measurements for core masses from 0 to 20 \me. These one-dimensional interior models include an inert core composed of either 50\% rock and 50\% ice or 100\% water ice. Each model has a homogenous convective envelope made of a H/He-rock-ice or H/He-ice mixture. We use equations of state from \ct{ANEOS} (water and water-rock) and \ct{SC95} (H/He), and atmospheric models from \ct{Fortney07a}. Neither the pure water or water-rock equations of state are dependent on temperature; since GJ 436b has a hot interior, we expect that the true equation of state of its interior will be more accurately matched by the water ice equation of state, even though the composition is likely a mix of heavy elements. 

We find that for the best-fit \tint\ of 336 K the maximum envelope metallicity is somewhat lower than 1000$\times$ solar metallicity. For a 20 \me\ core, the minimum envelope metallicity required to match the observed mass and radius is $\sim$300$\times$ solar metallicity (though the envelope could be less enriched if the core is somewhat larger than 20 \me. These calculations suggest that the very high metallicities above 1000$\times$ solar metallicity explored by our retrieval models are not physically realistic for this planet, but metallicities above 300$\times$ solar are favored.

Very high metallicities are only possible if accretion and subsequent enrichment is dominated by rocky rather than icy materials; \ct{Fortney13} show that if the majority of accretion is from icy material, the hydrogen in those ices is also accreted and the maximum metal-enrichment is about $\sim$600$\times$ solar metallicity. Though we cannot currently distinguish between compositions less than or greater than 600$\times$ solar composition, if GJ 436b is indeed very metal-enhanced, it likely formed in a region with more refractory than volatile materials available.

\subsection{Role of JWST Spectral Observations}

\jwst\ will amplify our understanding of warm Neptunes like GJ 436b by providing spectra instead of photometry, breaking some of the current degeneracies. For example, examining the spectra in Figure \ref{spectra_retrievals}, it is clear that models with very different spectra can have very similar photometry. \jwst\ may also allow us to detect molecules that are not currently included in most models; for example, \ct{Shabram11} showed that if species such as C$_2$H$_2$ and HCN exist in the atmosphere of GJ 436b, their abundances could be constrained by measuring the widths of features at 1.5, 3.3, and 7 \micron. 

\ct{Greene16} quantify our ability to constrain planet properties of a wider variety of atmospheres including hot Jupiters, warm Neptunes, warm sub-Neptunes, and cool super Earths with \jwst\ and find that the mixing ratios of major species in warm Neptunes like GJ 436b can be constrained to within better than 1 dex with a single secondary eclipse observation for each wavelength region from 1--11 \micron.

\subsection{Measuring Internal Dissipation Factor Using \tint}

Measuring \tint\ of GJ 436b using atmospheric models allows us to approximate the dissipation factor in GJ 436b's interior, $Q'$. $Q'$ is defined as $3Q/2k_2$, where $Q$ is the quality factor and $k_2$ is the Love number of degree 2 \cp{Goldreich66}. Our best-fit \tint\ from the retrieval analysis is 336 K. \ct{Agundez14} calculated relations between \tint\ and $Q'$ assuming obliquities of 0 and 15 degrees and 3 different rotation speeds (1:1 resonance, 3:2 resonance, and pseudo-synchronous). Assuming \tint$\sim$300--350 K, their calculations suggest that $Q'\sim 2\times10^5$--$10^6$. These values are somewhat larger than the value of $Q'$ that has been measured using Neptune's satellites of between 3.3$\times10^4$ and 1.35$\times10^5$ \cp{Zhang08}.  

If this high value for $Q'$ is correct, this has significant implications for both the structure of GJ 436b itself and the evolution of the planetary system. In particular, a high $Q'$ is consistent with a tidal circularization timescale that is longer than the age of the system, allowing GJ 436b to maintain its non-zero eccentricity ($e\sim$0.15) without invoking another object in the system \cp{Jackson08, Batygin09}. This is consistent with the observations to date that have not found a third body in the GJ 436 system despite extensive searches (see Section \ref{thirdbody}). If GJ 436b's $Q$ does differ significantly from Neptune's, this potentially implies a structural difference between hot Neptunes on short orbits and the ice giants in our own solar system.

\subsection{Condensation of graphite}

As has been discussed in, e.g., \ct{Moses13}, cool high metallicity atmospheres may have regions that are stable for the condensation of graphite. Indeed, the very high metallicity models favored by the retrieval models do indeed cross the graphite stability curve above 0.1 bar. While the effect of this condensation is beyond the scope of this work, the major effects would be twofold. First, the graphite condensation will deplete the carbon reservoir, decreasing the CO abundance in the upper atmosphere. In addition, the condensed graphite may form into cloud particles with their own opacity. Like other clouds and hazes, graphite clouds would likely decrease the size of features in transmission spectra and thermal emission spectra, and may either increase or decrease the albedo depending on the optical properties of the graphite particles.

\subsection{Spatially Inhomogeneous Clouds}

Planetary atmospheres are by their nature three dimensional and complex, and clouds in these atmospheres may be non uniformly located. In particular, the terminators of a planet have different circulation patterns and temperatures than the substellar point \cp[e.g.,][]{Lewis10, Kataria16}. It has recently been shown that non-uniform clouds on the terminators of planets will affect interpretation of transmission spectra \cp{Line16}. Furthermore, fitting the same one-dimensional model to both the thermal emission and transmission spectra may not be an accurate assumption. For example, the terminators may be cloudy while the thermal emission is dominated by a relatively cloud-free region. This effect should be investigated in the future, especially with higher resolution and signal-to-noise thermal emission and transmission spectra from \emph{JWST}.

\section{Conclusion}

We have presented new observations of GJ 436b's thermal emission at 3.6 and 4.5 \micron, which are in agreement with previous analyses from \ct{Lanotte14} and reduce the uncertainties of GJ 436b's flux at those wavelengths. For the first time, we combine these revised data with \emph{Spitzer} photometry from 5.6 to 16 \micron\ and transmission spectra from \emph{HST}/WFC3 and compare these data to both self-consistent and retrieval models. We vary the metallicity, internal temperature from tidal heating, disequilibrium chemistry, heat redistribution, and cloud properties. We find that our nominal best-fitting self-consistent model has 1000$\times$ solar metallicity, T$_{\rm int}$=240 K, \fsed=0.3 sulfide/salt clouds, disequilibrium chemistry, and planet-wide average temperature profile, but this model does not provide an accurate fit to the observations. Retrieval models find a statistically better fit to the ensemble data than the self-consistent model, with parameters in general agreement with the self-consistent approach: \emph{all signs point to a high metallicity, with best fits above several hundred times solar metallicity, and tidal heating warming its interior, with best-fit \tint$\sim$300--350 K.} These results are consistent with results from interior models to match the mass and radius, with core masses around 10 \me. 

While Neptune has been measured based on its methane abundance to have an atmospheric carbon enhancement of $\sim$100$\times$ solar, repeated observations of both the thermal emission and transmission spectra of the first exo-Neptune to be studied in detail, GJ 436b, have demonstrated that it likely has a significantly higher metallicity. Neptune itself may actually be more enhanced in other elements than it is in carbon; \ct{Luszcz-Cook13} infer a 400--600$\times$ solar enhancement in oxygen from microwave observations of upwelled CO in Neptune, though this cannot be verified with infrared spectra since oxygen is frozen into clouds. Studies of warmer exoplanet atmospheres will allow us to spectroscopically measure abundances of these molecules like oxygen that are locked into clouds in the cold ice giants of our solar system, potentially revealing unexpected patterns in the metal-enrichments of these intermediate-mass objects. 

An interesting new paradigm for this class of intermediate-sized planet is now being pieced together: we suggest that Neptune-mass planets may be more compositionally diverse than previously imagined. High quality data across of range of Neptune-mass planets with different temperatures and host stars will be critical to investigate the diversity of this class of planets. 

\acknowledgements We thank the anonymous referee for their helpful suggestions. We also thank Konstantin Batygin and Greg Laughlin for helpful discussions that improved the paper. This work was performed in part under contract with the Jet Propulsion Laboratory (JPL) funded by NASA through the Sagan Fellowship Program executed by the NASA Exoplanet Science Institute. JJF acknowledges Hubble grants HST-GO-13501.06-A and HST-GO-13665.004-A and NSF grant AST-1312545. MSM acknowledges support of the NASA Origins program. MRL acknowledges support provided by NASA through Hubble Fellowship grant \#51362 awarded by the Space Telescope Science Institute, which is operated by the Association of Universities for Research in Astronomy, In., for NASA, under the contract NAS 5-26555. This work is based on observations made with the \textit{Spitzer Space Telescope}, which is operated by the Jet Propulsion Laboratory, California Institute of Technology, under contract with NASA.


\end{document}